\documentclass[a4paper,twocolumn,11pt,accepted=2024-04-20]{quantumarticle}
\pdfoutput=1
\usepackage[utf8]{inputenc}
\usepackage[english]{babel}
\usepackage[T1]{fontenc}
\usepackage{amsmath}
\usepackage{hyperref}
\usepackage[numbers,sort&compress]{natbib}
\usepackage{tikz}
\usepackage{lipsum}
\usepackage{subfigure}

\begin{document}
	
	\title{Kerr-effect-based quantum logical gates in decoherence-free subspace}
	
	\author{Fang-Fang Du*}
	\email{dufangfang19871210@163.com}
	\orcid{0000-0002-1213-3041}
	
	\author{Gang Fan}
	\email{fg18235091403@163.com}
	\orcid{0000-0002-8521-263X}

	\author{Xue-Mei Ren}
	
	\affiliation{Science and Technology on Electronic Test
		and Measurement Laboratory, North University of China, Taiyuan 030051, China}

	\maketitle
	
	\begin{abstract}
		The decoherence effect caused by the coupling between the system and the environment undoubtedly leads to the errors in efficient implementations of two (or three) qubit logical gates in quantum information processing. Fortunately, decoherence-free subspace (DFS) introduced can effectively decrease the influence of decoherence effect.
		In this paper, we propose some schemes for setting up a family of quantum control gates, including controlled-NOT (CNOT),  Toffoli, and Fredkin gates for two or three logical qubits by means of cross-Kerr nonlinearities in DFS.
		These three logical gates
		require neither complicated quantum computational circuits nor auxiliary photons (or entangled states).
		The success probabilities of three logical gates are approximate 1
		by performing the corresponding classical feed-forward operations
		based on the different measuring results of the X-homodyne detectors, and their fidelities are robust
		against the photon loss with the current technology.
		The proposed logical gates rely on only simple linear-optics elements, available single-qubit operations, and mature measurement methods, making our proposed gates be feasible and efficient in practical applications.
	\end{abstract}
	

	\section{Introduction}\label{sec1}
	
	Quantum computing, as a very sought-after technology, is used to tackle problems unsolvable by the best-of-breed classical computing \cite{Quantum2002,Quan1},
	which offers remarkable benefits, such as tremendous speedup \cite{Quan2}, efficient searching for unordered databases \cite{Quan3}, large number factorization \cite{Quan4}, and so on.
	Further, quantum computing is an imperative component of useful quantum information processing (QIP),
	making it ideal for adhibition in QIP \cite{long2002theoretically,zhang2017quantum,zhu2017experimental,du2019efficient,li2020quantum,Quan5,WANG202191,qi202115,long2021drastic,sheng2022one,2024Cao-oe, exper, PhysRevA.105.032609}. The foundation of quantum computing is rest with quantum logical gates, which connect qubits and allow for applications of large-scale realization of quantum computation. The universal quantum gates \cite{cao2019high}, e.g., the two-qubit CNOT (controlled-NOT), three-qubit Toffoli (controlled-controlled-NOT), and Fredkin (controlled-swap) gates, can simplify complex quantum circuits to achieve the goals of QIP \cite{CaoBELL,wu2022FOP,Zhang_2023} with various quantum information platforms,
	such as linear optics \cite{Wei-npj,Wei-PRApplied,17,37,wei2024PRApplied}, cross-Kerr nonlinearity \cite{KerrCNOT,KerrCNOT1,KerrCNOT2,Tgate-Dong,adv2}, nitrogen-vacancy centers \cite{weiNV,PRB-AiQing,Rxm_OL}, quantum dots \cite{WeiQD,wei-cpf,cao}, waveguide systems \cite{SongPRA,LiPRA1,SongPRA1}, and neutral atoms \cite{DUcnot,adv1}.
	
	Nowadays, Google has alleged the quantum
	supremacy \cite{Google}, unfortunately, the superconducting qubit processors
	are subjected to the loss of quantum coherence and the difficulty
	of scalability originated from expensive cryogenic apparatus
	strongly coupling to the realistic environment. That is, photon technologies have served relatively economical
	solutions, and photon-encoded qubits are also agreed with both
	low-decoherence applications and carriers to transmit the QIP.
	So far,
	some important progresses
	have been achieved in demonstrating the functionality
	of quantum logic gates for photon systems \cite{17,37,Wei-npj,Wei-PRApplied}.
	Additionally,
	the polarization degrees of freedom of photon system is easy to encode and manipulate with current optics techniques.
	However, the interference of environmental factors (i.e., atmospheric fluctuation, thermal and mechanical fluctuation, and birefringence in optical fibers) can lead to undesired interactions (harmful decoherence) in practical circumstances. These factors can cause errors in the transmission of polarized states of photons, making it necessary to find useful ways to sidestep their impact on QIP. 
	The decoherence-free subspace (DFS) approach has garnered significant attention to overcome the adverse effects of environmental noise factors on quantum information transmission,
	where DFS is a subspace of all quantum states in the unitary evolution of the quantum system.
	By utilizing two photons to encode a single logic qubit in DFS, i.e.,
	$|\overline{0}\rangle=|H\rangle_{1}|V\rangle_{2}=|HV\rangle$ and $|\overline{1}\rangle=|V\rangle_{1}|H\rangle_{2}=|VH\rangle$ ($|H\rangle$ and $|V\rangle$ denote horizontally and vertically polarized states of photons, respectively) in Ref. \cite{HV-encode}.
	After passing through a noisy channel, the state of the one photon is changed into $|H\rangle\xrightarrow[]{ transmission}e^{i\gamma_{1}}|H\rangle$
	and $|V\rangle\xrightarrow[]{ transmission}e^{i\gamma_{2}}|V\rangle$. Then the evolution of the superposition state in the decoherence
	environment can be expressed as
	\begin{eqnarray}  
		&&\quad\alpha e^{i\gamma_{1}}|H\rangle+\beta e^{i\gamma_{2}}|V\rangle \nonumber \\
		&&=e^{i\gamma_{1}}(\alpha|H\rangle +\beta e^{i(\gamma_{2}-\gamma_{1})}|V\rangle)  \nonumber \\
		&&=\alpha|H\rangle +\beta e^{i\gamma}|V\rangle,
	\end{eqnarray}
	where $\gamma=\gamma_{2}-\gamma_{1}$. Afterward, the evolution of the state in DFS can be illustrated as
	\begin{eqnarray}   
		\alpha|\overline{0}\rangle+\beta|\overline{1}\rangle&=&\alpha|H\rangle_{1}|V\rangle_{2}+\beta|V\rangle_{1}|H\rangle_{2}   \nonumber \\
		&\rightarrow&\alpha|H\rangle_{1}e^{i\gamma}|V\rangle_{2}+\beta e^{i\gamma}|V\rangle_{1}|H\rangle_{2}  \nonumber \\
		&=&e^{i\gamma}(\alpha|H\rangle_{1}|V\rangle_{2}+\beta|V\rangle_{1}|H\rangle_{2})  \nonumber \\
		&=&\alpha|\overline{0}\rangle+\beta|\overline{1}\rangle.
	\end{eqnarray}
	Obviously, it is noted that DFS can effectively avoid the influence of collective dephasing noise. Therefore, quantum states encoded on the DFS are robust against certain types of noise and decoherence processes. This resilience makes DFS to have applications in various areas of QIP, including the construction of the quantum network \cite{QED-DFS}, the valid preparation and the conversion of the quantum states \cite{entangled-states, KLM, GHZ,Rxm_OE,  WOS:000766964200004}, and quantum teleportation \cite{teleportation, WOS:000501539900004}, quantum computation \cite{PhysRevA.104.062612, PhysRevLett.85.1758, chen2010quantum}, and quantum sensing \cite{PhysRevLett.125.090501, hamann2022approximate}. They offer a promising avenue for mitigating the effects of decoherence and improving the reliability of quantum technologies.

	In this paper, in view of the cross-Kerr nonlinearity, we present some schemes for setting up a family of quantum control gates, including CNOT,  Toffoli, and Fredkin gates in DFS for polarization-encoded logical qubits of photon systems, which has the advantages over low-decoherence character, flexible single-qubit manipulation, and ultra-fast transmission.
	The employment of two key tools, i.e., a SWAP gate (abbreviated as S) and a path coupler, are innovated for implementing three logical gates.
	The former with the success probability of 1 relies on only simple linear-optics elements, which swaps not only the polarized states of two photons but also two
	paths of a single photon.
	The latter with the probability near close to 1 is utilized to combine two paths of a photon to the one by available single-qubit operations and mature measurement methods, which halves the number of the paths of the photon and greatly simplifies the quantum circuit.
	Therefore,
	these three logical gates
	require neither complicated quantum computational circuits nor auxiliary photons (or entangled states).
	The success probabilities of three logical gates are approximate 1
	by performing the corresponding classical feed-forward operations
	based on the different measuring results of the X-homodyne detectors, and their fidelities are robust
	against the photon loss with the current technology.


	\section{\label{sec:2} SWAP gate and path coupler }
	
	Nowadays, to break up the limitation on the success probabilities less
	than 1 of the quantum gates with linear optics \cite{17,37,Wei-npj,Wei-PRApplied}, the
	cross-Kerr nonlinearity \cite{KerrCNOT,KerrCNOT1,KerrCNOT2,parity-gate,Tgate-Dong}
	has provided a potential solution for improving the success probability of the quantum gate,
	as it has the potential available to an
	evolution of the combined system composed of $n$ photons and a coherent state but without destroying
	the involved photon.
	Let us assume that $|n\rangle_{i}$ represents a state of n photons (signal), and the probe beam is in a coherent state $|\alpha\rangle_{P}$. After passing through the interaction $U_{K}$ of cross-Kerr nonlinearity,
	the conditional phase shift between a signal system and probe beam as follows:
	\begin{eqnarray}  \label{eq41}
		U_{K}|n\rangle_{i}|\alpha\rangle_{P}=e^{-i\frac{t}{\hbar}\hat{H}}|n\rangle_{1}|\alpha\rangle_{P}=|n\rangle_{1}|\alpha e^{in\theta}\rangle,
	\end{eqnarray}
	where $\hat{H}=\hbar\chi N_{1}N_{2}$ is the interaction Hamiltonian of cross-Kerr nonlinearity. The Kerr effect varies with different media and $\chi$ represents the strength of nonlinearity.
	$N_{1}$ and $N_{2}$ are photon number operators of signal and probe modes, respectively. $\theta=\chi t$ is the phase shift and $t$ is the interaction time in Kerr medium. $|n\rangle$ is the $n$-photon state, and $|\alpha\rangle=e^{-|\alpha|^{2}/2}\sum\limits^{\infty}_{n=0}[\alpha^{n}/\sqrt{n!}]|n\rangle$ is the coherent state of probe field.

	\begin{figure}[t]
		\includegraphics[width=8cm,angle=0]{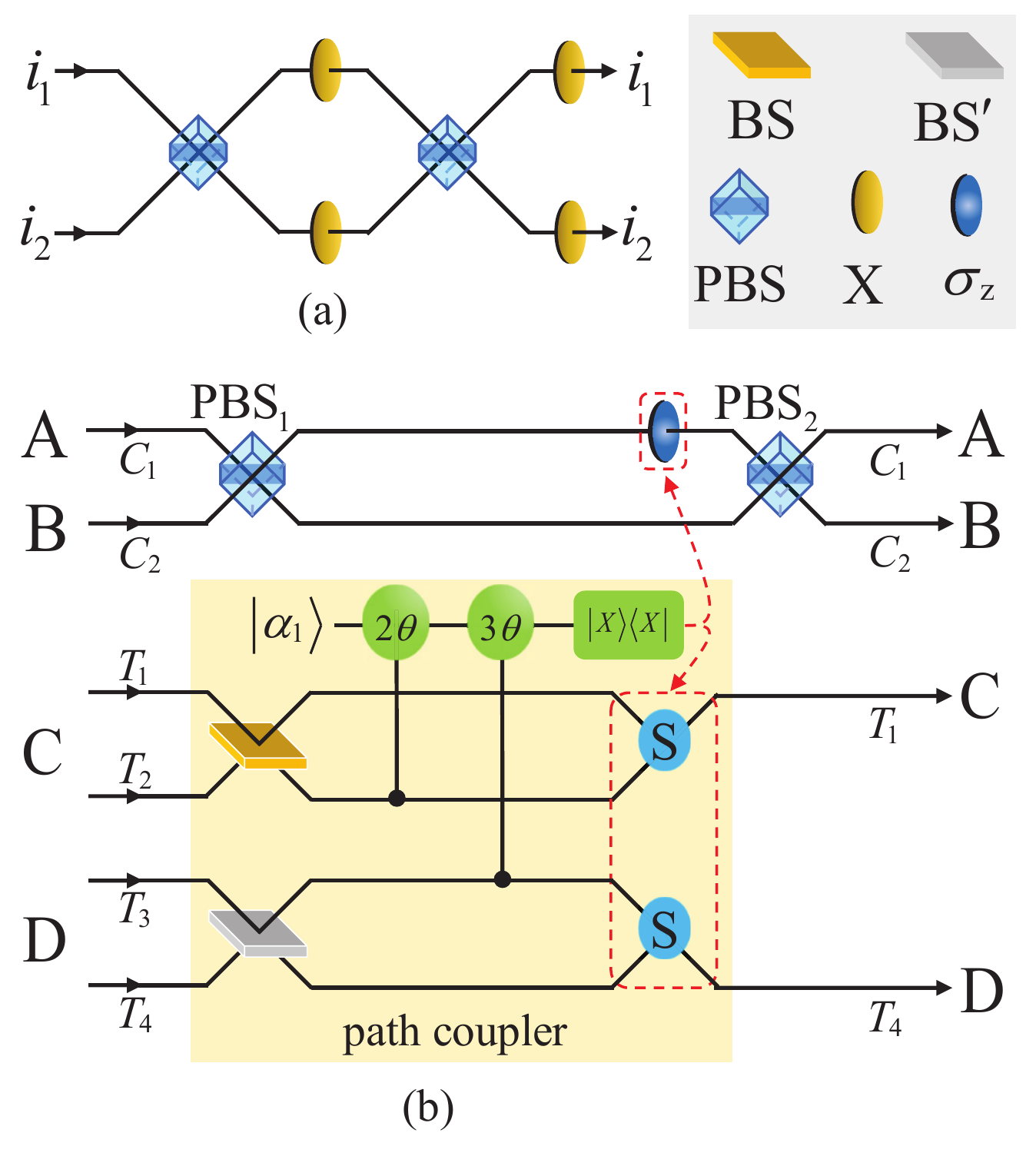}
		\caption{\label{fig1}(a) Schematic diagram of a SWAP gate (abbreviated as S) swapping the polarized states of two photons and  two paths of a single photon as well.
			(b) Schematic diagram of a path coupler to combine two paths into one without destroying the polarization information.
			BS represents a $50:50$ beam splitter, which completes the paths transformation between up and down paths $(i_{up}, i_{down})$, $i_{up}\rightarrow(i_{up}+i_{down})/\sqrt{2}, i_{down}\rightarrow(i_{up}-i_{down})/\sqrt{2}$, while the other BS$'$ completes the transformation $i_{up}\rightarrow(i_{down}-i_{up})/\sqrt{2}, i_{down}\rightarrow(i_{up}+i_{down})/\sqrt{2}$.
			Polarization beam splitters (PBSs) transmit the $|H\rangle$-polarized component and reflect the $|V\rangle$-polarized one of the photon.
			The X  fulfills the bit-flip operation $|H\rangle\leftrightarrow|V\rangle$ and the $\sigma_{z}$  fulfills the phase-flip operation $\sigma_{z}=|H\rangle\langle H|-|V\rangle\langle V|$.}
	\end{figure}

	\subsection{SWAP gate}\label{sec:2-1}

	The SWAP gate (abbreviated as S), depicted in Figure \ref{fig1}a, serves two functions. The one is to exchange two paths of a single photon and the other one is to swap the polarized states of two photons. In detail,
	suppose that the input photon 1 and  photon 2 are initially in the following state
	\begin{eqnarray}  \label{eq1}
		\!\!\!\!\!|S\rangle_{0}&=&
		(\xi_{1}|H\rangle+\xi_{2}|V\rangle)_{1}|i_{1}\rangle \nonumber \\
		&&\otimes
		(\epsilon_{1}|H\rangle+\epsilon_{2}|V\rangle)_{2}|i_{2}\rangle,
	\end{eqnarray}
	where $i_{1}$ and $i_{2}$ represent the paths of input photon 1 and photon 2, respectively. The complex coefficients conform to the normalization principle, i.e., $|\xi_{1}|^{2}+|\xi_{2}|^{2}=1$ and $|\epsilon_{1}|^{2}+|\epsilon_{2}|^{2}=1$.
	After the input photon 1 and photon 2 pass through the optical elements polarization beam splitters (PBSs), which transmit the $|H\rangle$-polarized component and reflect the $|V\rangle$-polarized one of the photon, and Xs, where half wave plates oriented at 45$^{\circ}$ can fulfill the bit-flip operation $|H\rangle\leftrightarrow|V\rangle$, the parallel transformation can be obtained
	\begin{eqnarray} \label{eq2}
		\!\!\!\!\!|S\rangle_{1}&=&
		(\epsilon_{1}|H\rangle+\epsilon_{2}|V\rangle)_{2}|i_{1}\rangle \nonumber \\
		&&\otimes(\xi_{1}|H\rangle+\xi_{2}|V\rangle)_{1}|i_{2}\rangle.
	\end{eqnarray}
	Based on Eq. (\ref{eq2}), the quantum circuit unequivocally swaps the polarized states of two photons 12 only with simple linear-optics elements.
	Additionally, the S has another function, which can swap two paths of a single photon, as the single photon enters from the up (down) path  $|i_{1}\rangle$ ($|i_{2}\rangle$) and then
	exits from the down (up) one $|i_{2}\rangle$ ($|i_{1}\rangle$).

	\subsection{ Path coupler}\label{sec:2-2}

	
	A path coupler shown in Figure \ref{fig1}b, called a quantum eraser, 
	combines two paths into one,
	making halve the number of the paths without destroying the polarization information of the photon.
	Suppose that
	the state of two photons AB to encode a logic qubit and the state of two photons CD to encode another logic qubit in DFS are
	\begin{eqnarray}  \label{eq4}
		|\phi\rangle_{AB}&=&\beta_{1}|\overline{0}\rangle+\beta_{2}|\overline{1}\rangle, \nonumber\\
		|\phi\rangle_{CD}&=&\delta_{1}|\overline{0}\rangle+\delta_{2}|\overline{1}\rangle,
	\end{eqnarray}
	where the complex coefficients conforms to the normalization principle, i.e., $|j_{1}|^{2}+|j_{2}|^{2}=1, (j = \beta, \delta)$.
	Four photons ABCD are initially prepared in the arbitrary state
	\begin{eqnarray} \label{eq5}
		|\Gamma\rangle_{0}
		&=&\beta_{1}|\overline{0}\rangle_{AB}|\phi\rangle_{CD}|C_{1}C_{2}\rangle|T_{1}T_{4}\rangle   \nonumber \\
		&&+\beta_{2}|\overline{1}\rangle_{AB}|\phi\rangle_{CD}|C_{1}C_{2}\rangle|T_{2}T_{3}\rangle.
	\end{eqnarray}
	Here, ($C_{1}, C_{2}$) represent the paths of photon AB. ($T_{1}$, $T_{2}$) and ($T_{3}$, $T_{4}$) represent the paths of photon C and photon D, respectively.

	Firstly, photons AB pass through $\rm PBS_{1}$,  meanwhile photons CD pass through $50:50$ beam splitters, i.e., BS and BS$'$,  where BS completes the paths transformation between up path $i_{up}$ and down path $ i_{down}$: $i_{up}\rightarrow(i_{up}+i_{down})/\sqrt{2}, i_{down}\rightarrow(i_{up}-i_{down})/\sqrt{2}$, while BS$'$ completes the transformation: $i_{up}\rightarrow(i_{down}-i_{up})/\sqrt{2}, i_{down}\rightarrow(i_{up}+i_{down})/\sqrt{2}$), respectively. The system state is changed into
	\begin{eqnarray}
		|\Gamma\rangle_{0}'
		&=&\beta_{1}|\overline{0}\rangle_{AB}|\phi\rangle_{CD}|C_{1}C_{2}\rangle   \nonumber \\
		&&\otimes(|T_{1}T_{3}\rangle+|T_{1}T_{4}\rangle+|T_{2}T_{3}\rangle+|T_{2}T_{4}\rangle)   \nonumber \\
		&&+\beta_{2}|\overline{1}\rangle_{AB}|\phi\rangle_{CD}|C_{1}C_{2}\rangle   \nonumber \\
		&&\otimes(|T_{1}T_{4}\rangle-|T_{1}T_{3}\rangle-|T_{2}T_{4}\rangle+|T_{2}T_{3}\rangle).\nonumber \\
	\end{eqnarray}
	Then the photons CD interact with the coherent state $|\alpha\rangle_{1}$ via Kerr media, based on Eq. (\ref{eq41}), the process can be expressed as
	\begin{eqnarray}
		|\Gamma\rangle_{1}&=&
		\frac{1}{2}(\beta_{1}|\overline{0}\rangle_{AB}|\phi\rangle_{CD}|C_{2}\rangle|T_{1}T_{4}\rangle   \nonumber \\
		&&+\beta_{2}|\overline{1}\rangle_{AB}|\phi\rangle_{CD}|C_{1}\rangle|T_{1}T_{4}\rangle)|\alpha_{1}\rangle   \nonumber\\
		&&+(\beta_{1}|\overline{0}\rangle_{AB}|\phi\rangle_{CD}|C_{2}\rangle|T_{2}T_{4}\rangle   \nonumber \\
		&&-\beta_{2}|\overline{1}\rangle_{AB}|\phi\rangle_{CD}|C_{1}\rangle|T_{2}T_{4}\rangle)|\alpha_{1}e^{2i\theta}\rangle    \nonumber\\
		&&+(\beta_{1}|\overline{0}\rangle_{AB}|\phi\rangle_{CD}|C_{2}\rangle|T_{1}T_{3}\rangle   \nonumber \\
		&&-\beta_{2}|\overline{1}\rangle_{AB}|\phi\rangle_{CD}|C_{1}\rangle|T_{1}T_{3}\rangle)|\alpha_{1}e^{3i\theta}\rangle    \nonumber\\
		&&+(\beta_{1}|\overline{0}\rangle_{AB}|\phi\rangle_{CD}|C_{2}\rangle|T_{2}T_{3}\rangle   \nonumber \\
		&&+\beta_{2}|\overline{1}\rangle_{AB}|\phi\rangle_{CD}|C_{1}\rangle|T_{2}T_{3}\rangle)|\alpha_{1}e^{5i\theta}\rangle.
	\end{eqnarray}
	After following the X-homodyne measurement,  the interaction of the cross-Kerr nonlinearity can induce 
	four measurement outcomes, i.e., phase shifts $0$, $2\theta$, $3\theta$ and $5\theta$.
	If the outcome is zero phase shift, the state of the whole system becomes into
	\begin{eqnarray}  \label{eq8}
		|\Gamma\rangle_{2}&=&
		(\beta_{1}|\overline{0}\rangle_{AB}|C_{2}\rangle+\beta_{2}|\overline{1}\rangle_{AB}|C_{1}\rangle)\nonumber \\
		&&\otimes|\phi\rangle_{CD}|T_{1}T_{4}\rangle.
	\end{eqnarray}
	If the outcome is another case, corresponding feed-forward operations shown in Table \ref{Table:1} must be performed based on the measurement outcomes to get the desirous state in Eq. (\ref{eq8}). In Table \ref{Table:1}, $I_{T_{1}, T_{2}}$ ($I_{T_{3}, T_{4}}$) represents the paths $T_{1}, T_{2}$ $(T_{3}, T_{4})$ of photons C (or D) to remain unchanged, $S_{T_{1}, T_{2}}$ ($S_{T_{3}, T_{4}}$) mean to swap two paths of the photon C (or D) with the S in Figure \ref{fig1}a, and $\sigma_{z}$ performs the phase-flip operation $\sigma_{Z}=|H\rangle\langle H|-|V\rangle\langle V|$ from the path $C_{1}$.

	\begin{table}[htbp]
		\caption{\label{Table:1} The measurement results of the coherent state $|\alpha_{1}\rangle$, and the corresponding single-qubit feed-forward operations.}
		\begin{tabular}{cc} \colrule
			\multicolumn{1}{c}{\begin{tabular}[c]{@{}c@{}}Measurement results\end{tabular}}&    \multicolumn{1}{c}{Single-qubit operations}\\
			\hline
			0        &      \multicolumn{1}{c}{$I_{T_{1},T_{2}}\otimes I_{T_{3},T_{4}}$}               \\
			$2\theta$            &      \multicolumn{1}{c}{$(\sigma_{z})_{C_{1}}\otimes I_{T_{1},T_{2}}\otimes S_{T_{3},T_{4}}$}    \\
			$3\theta$           &      \multicolumn{1}{c}{$(\sigma_{z})_{C_{1}}\otimes S_{T_{1},T_{2}}\otimes I_{T_{3},T_{4}}$}   \\ 
			$5\theta$          &         \multicolumn{1}{c}{$S_{T_{1},T_{2}}\otimes S_{T_{3},T_{4}}$}           \\ \hline
		\end{tabular}
	\end{table}

	Finally, photons AB pass through the $\rm PBS_{2}$, resulting in the final state
	\begin{eqnarray}
		|\Gamma\rangle_{3}\!&=&\!
		(\beta_{1}|\overline{0}\rangle+\beta_{2}|\overline{1}\rangle)_{AB}|\phi\rangle_{CD}|C_{1}C_{2}\rangle|T_{1}T_{4}\rangle. \nonumber \\
	\end{eqnarray}
	That is, the path coupler merges two paths $T_{1}$ and $T_{2}$ ($T_{3}$ and $T_{4}$) into the one $T_{1}$ ($T_{4}$) with the above device, and likewise, as the quantum circuit in Figure \ref{fig1}a is universal and flexible,
	the merging of other paths we desired can be obtained by suitable feed-forward operations, such as $T_{1}$ and $T_{2}$ ($T_{3}$ and $T_{4}$) are merged into the path $T_{2}$ ($T_{3}$).
	Similarly, when the initial state of two photons CD to encode logic qubit in DFS is
	$\overline{|\phi\rangle}_{CD}=\delta_{1}|\overline{1}\rangle+\delta_{2}|\overline{0}\rangle$,
	by replacing $|\phi\rangle_{CD}$ with $\overline{|\phi\rangle}_{CD}$ in Eq. (\ref{eq5}), we also obtain the same appearance.
	That is, whether the initial state is $|\phi\rangle_{CD}$ or $\overline{|\phi\rangle}_{CD}$,
	the path coupler can merges two paths into one.

	\section{Quantum logical gate}\label{sec:3}
	
	\subsection{CNOT gate}\label{sec:3.1}
	
	\begin{figure}[t]
		\includegraphics[width=8cm,angle=0]{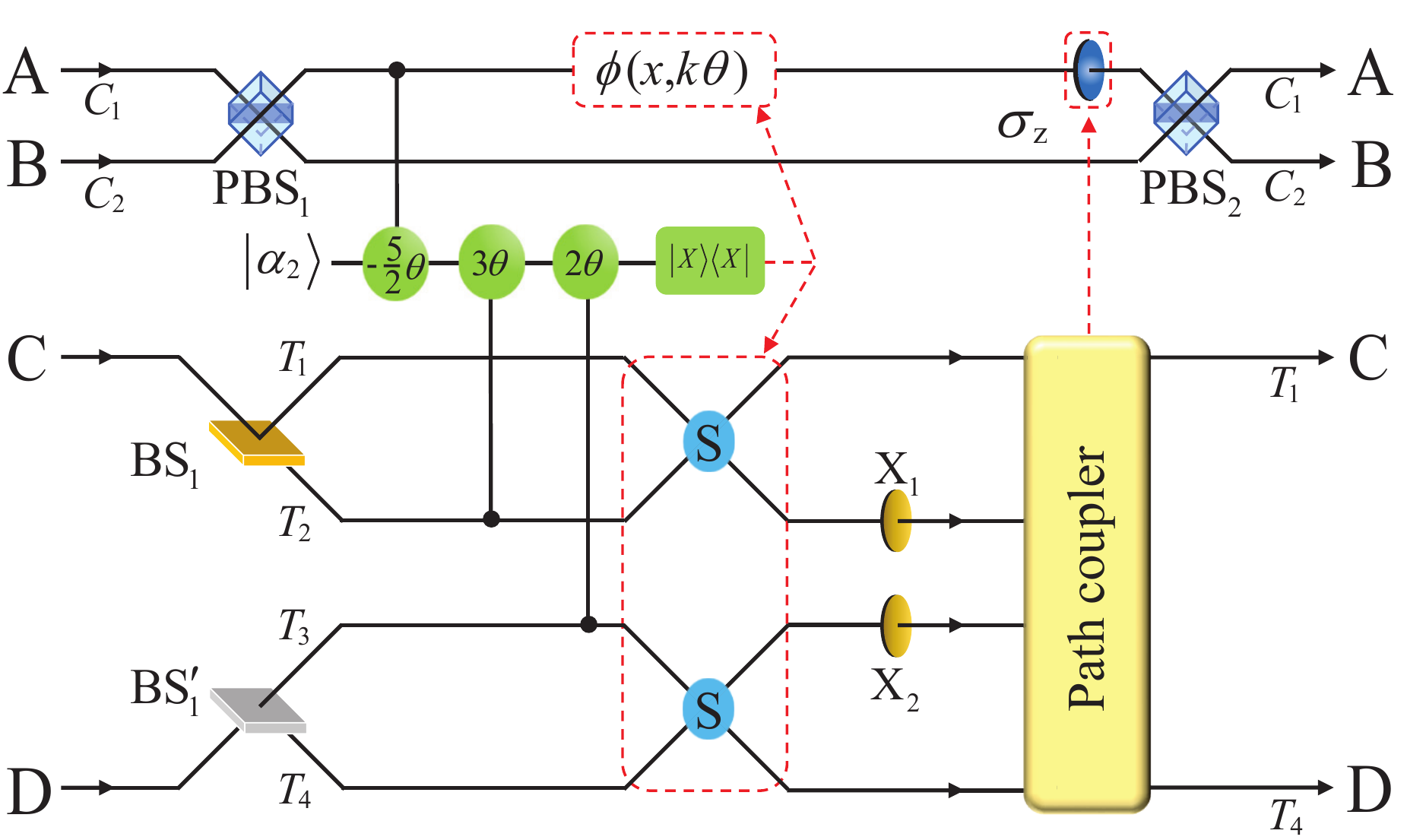}
		\caption{\label{fig2} Schematic diagram of the CNOT gate based on cross-Kerr nonlinearities in DFS.
			$\phi(x,k\theta)(k= 2, 3, 5)$ represents the phase modulation.}
	\end{figure}
	
	Suppose that the initial state of two logical qubits in DFS is $|\Omega\rangle_{0}=|\phi\rangle_{AB}\otimes |\phi\rangle_{CD}$, where the quantum state $|\phi\rangle_{AB}$ of control logical qubit AB and that $|\phi\rangle_{CD} $ of target logical qubit CD are the same as those in Eq. (\ref{eq4}).
	The quantum circuit for implementing the CNOT gate is shown in Figure \ref{fig2}. Firstly, the control photons  AB pass through $\rm PBS_{1}$, meanwhile the target photons CD pass through $\rm BS_{1}$ and $\rm BS_{1}'$, respectively. Then, four photons combine with the coherent state $|\alpha\rangle_{2}$ with the help of the cross-Kerr nonlinearity, resulting in the quantum state $|\Omega\rangle_{0}=|\phi\rangle_{AB}\otimes |\phi\rangle_{CD}\otimes|\alpha_{2}\rangle$  changed into the state
	\begin{eqnarray}   \label{eq9}
		|\Omega\rangle_{1}&\!\!=\!\!&
		\frac{1}{2}[(\beta_{1}|HV\rangle_{AB}|\phi\rangle_{CD}|C_{2}\rangle|T_{1}T_{4}\rangle  \nonumber\\
		&&\!\!+\beta_{2}|VH\rangle_{AB}|\phi\rangle_{CD}|C_{1}\rangle|T_{2}T_{3}\rangle)|\alpha_{2}\rangle   \nonumber\\
		&&\!\!+(\beta_{1}|HV\rangle_{AB}|\phi\rangle_{CD}|C_{2}\rangle|T_{1}T_{3}\rangle|\alpha_{2}e^{2i\theta}\rangle  \nonumber\\
		&&\!\!+\beta_{2}|VH\rangle_{AB}|\phi\rangle_{CD}|C_{1}\rangle|T_{2}T_{4}\rangle|\alpha_{2}e^{-2i\theta}\rangle)  \nonumber\\
		&&\!\!+(\beta_{1}|HV\rangle_{AB}|\phi\rangle_{CD}|C_{2}\rangle|T_{2}T_{4}\rangle|\alpha_{2}e^{3i\theta}\rangle  \nonumber\\
		&&\!\!+\beta_{2}|VH\rangle_{AB}|\phi\rangle_{CD}|C_{1}\rangle|T_{1}T_{3}\rangle|\alpha_{2}e^{-3i\theta}\rangle)   \nonumber\\
		&&\!\!+(\beta_{1}|HV\rangle_{AB}|\phi\rangle_{CD}|C_{2}\rangle|T_{2}T_{3}\rangle|\alpha_{2}e^{5i\theta}\rangle  \nonumber\\
		&&\!\!+\beta_{2}|VH\rangle_{AB}|\phi\rangle_{CD}|C_{1}\rangle|T_{1}T_{4}\rangle|\alpha_{2}e^{-5i\theta}\rangle)].\nonumber \\
	\end{eqnarray}

	As the X-homodyne measurement cannot distinguish between phase shift $k\theta$ and phase shift $-k\theta$ $(k = 2, 3, 5)$, we obtain four different scenarios of phase shifts $(0, \pm2\theta, \pm3\theta, \pm5\theta)$.
	A phase modulation $\phi(x, k\theta)= \alpha\sin k\theta(x-2\alpha\cos k\theta)[\rm mod\; 2\pi]$ $(k = 2, 3, 5)$, a function of the phase shift and the eigenvalue $x$ of the X-homodyne operator, is performed to erase the phase difference between the two superposition terms in the scenarios of three different nonzero phase shifts.
	If the outcome is zero phase shift, the state of the whole system becomes into
	\begin{eqnarray}  \label{eq10}
		|\Omega\rangle_{1}'&=&\beta_{1}|HV\rangle_{AB}|\phi\rangle_{CD}|C_{2}\rangle|T_{1}T_{4}\rangle  \nonumber\\
		&&+\beta_{2}|VH\rangle_{AB}|\phi\rangle_{CD}|C_{1}\rangle|T_{2}T_{3}\rangle.
	\end{eqnarray}
	By performing corresponding feed-forward operations for the other measurement outcomes, as shown in Table \ref{Table:2},
	we can get the state $|\Omega\rangle_{1}$ with the success probability close to 1.

	Next, the target photons CD pass through $\rm X_{1}$ and $\rm X_{2}$ on paths $T_{2}$ and $T_{3}$, respectively, and the evolution of the system can be illustrated as follows
	\begin{eqnarray}
		|\Omega\rangle_{2}&=&\beta_{1}|\overline{0}\rangle_{AB}|\phi\rangle_{CD}|C_{2}\rangle|T_{1}T_{4}\rangle  \nonumber\\
		&&+\beta_{2}|\overline{1}\rangle_{AB}\overline{|\phi\rangle}_{CD}|C_{1}\rangle|T_{2}T_{3}\rangle.
	\end{eqnarray}
	
	\begin{table}[htbp]
		\caption{\label{Table:2} The measurement results of the coherent state $|\alpha_{2}\rangle$, and the corresponding single-qubit feed-forward operations.}
		\begin{tabular}{ccr} \hline
			\multicolumn{1}{c}{\begin{tabular}[c]{@{}c@{}}Measurement\\ results\end{tabular}}  &\multicolumn{1}{c}{\begin{tabular}[c]{@{}c@{}}Phase \\modulation\end{tabular}}  & \multicolumn{1}{c}{\begin{tabular}[c]{@{}c@{}}Single-qubit\\ operations\end{tabular}}\\
			\hline
			0       &0 &      \multicolumn{1}{c}{$I_{T_{1},T_{2}}\otimes I_{T_{3},T_{4}}$}               \\
			$\pm2\theta$           &\multicolumn{1}{c}{$\phi(x, 2\theta)_{C_{1}}$} &      \multicolumn{1}{c}{$I_{T_{1},T_{2}}\otimes S_{T_{3},T_{4}}$}    \\
			$\pm3\theta$          &\multicolumn{1}{c}{$\phi(x, 3\theta)_{C_{1}}$} &      \multicolumn{1}{c}{$S_{T_{1},T_{2}}\otimes I_{T_{3},T_{4}}$}   \\
			$\pm5\theta$          &\multicolumn{1}{c}{$\phi(x, 5\theta)_{C_{1}}$} &         \multicolumn{1}{c}{$S_{T_{1},T_{2}}\otimes S_{T_{3},T_{4}}$}            \\\hline
		\end{tabular}
	\end{table}
	
	Finally, the control photons AB pass through $\rm PBS_{2}$ and the target photons CD pass through the path coupler in Figure \ref{fig1}b, merging two paths $T_{1}$ and $T_{2}$ ($T_{3}$ and $T_{4}$) into one path $T_{1}$ ($T_{4}$) of the target photon C (D), the quantum state $|\Omega\rangle_{2}$ is evolved as
	\begin{eqnarray}
		|\Omega\rangle_{3}&=&(\beta_{1}|\overline{0}\rangle_{AB}|\phi\rangle_{CD}
		+\beta_{2}|\overline{1}\rangle_{AB}\overline{|\phi\rangle}_{CD}) \nonumber \\
		&&\otimes|C_{1}C_{2}\rangle|T_{1}T_{4}\rangle. 
	\end{eqnarray}
	It is obvious that the result of the CNOT gate flips the state of the target photons CD if and only if the control photons AB is in the state $|\overline{1}\rangle_{AB}$, and has no change otherwise.

	\subsection{Toffoli gate}\label{sec:3.2}
	
	\begin{figure*}[t]
		\centering
		\includegraphics[width=13cm,angle=0]{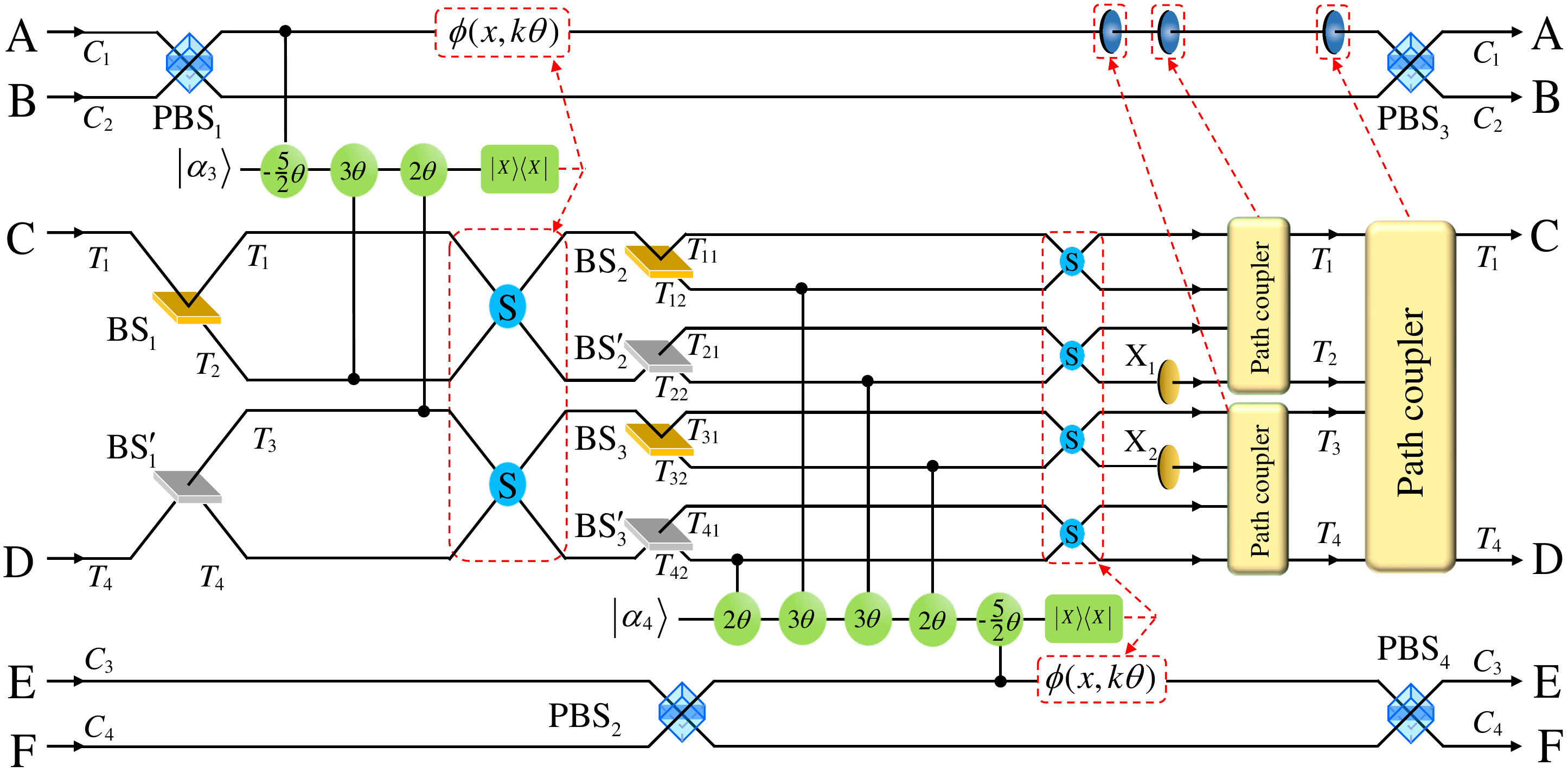}
		\caption{\label{fig3} Schematic diagram of the Toffoli gate in DFS based on cross-Kerr nonlinearities.
		}
	\end{figure*}
	
	The quantum circuit of the three-logic-qubit Toffoli gate composed of six photons in Figure \ref{fig3}, where the initial state of the second control qubit encoded by two photons EF in DFS is $|\phi\rangle_{EF}=\gamma_{1}|\overline{0}\rangle
	+\gamma_{2}|\overline{1}\rangle$ $(|\gamma_{1}|^{2}+|\gamma_{2}|^{2}=1)$.
	Firstly, the control photons AB and the target ones CD perform the same operations as the first step of two-logic-qubit CNOT gate in Figure \ref{fig2}.
	By performing corresponding feed-forward operations as shown in Table \ref{Table:2},
	we can get quantum state $|\Phi\rangle_{1}$ with the same form in Eq. (\ref{eq10}) with the success probability close to 1. Now, the system state can be expressed as $|\Phi\rangle_{1}=\beta_{1}|HV\rangle_{AB}|\phi\rangle_{EF}|\phi\rangle_{CD}|C_{2}\rangle|C_{3}C_{4}\rangle|T_{1}T_{4}\rangle+\beta_{2}|VH\rangle_{AB}|\phi\rangle_{EF}|\phi\rangle_{CD}|C_{1}\rangle|C_{3}C_{4}\rangle|T_{2}T_{3}\rangle$.
	
	Secondly, the second control photons EF enter into $\rm PBS_{2}$, meanwhile the target photons CD enter into ($\rm BS_{2}$, $\rm BS_{2}'$) and $\rm BS_{3}$ and $\rm BS_{3}'$, and their evolution interacting with the coherent state $\alpha_{4}$ via Kerr media can be expressed as
	\begin{widetext}
		\begin{eqnarray} 
			|\Phi\rangle_{2}  
			&=&(\beta_{1}\gamma_{1}|HV\rangle_{AB}|HV\rangle_{EF}|\phi\rangle_{CD}|C_{2}\rangle|C_{4}\rangle|T_{11}T_{41}\rangle  \nonumber\\  
			&&\!\!\!\!\!+\beta_{2}\gamma_{1}|VH\rangle_{AB}|HV\rangle_{EF}|\phi\rangle_{CD}|C_{1}\rangle|C_{4}\rangle|T_{21}T_{31}\rangle \nonumber\\	
			&&\!\!\!\!\!+\beta_{1}\gamma_{2}|HV\rangle_{AB}|VH\rangle_{EF}|\phi\rangle_{CD}|C_{2}\rangle|C_{3}\rangle|T_{12}T_{42}\rangle  \nonumber\\
			&&\!\!\!\!\!+\beta_{2}\gamma_{2}|VH\rangle_{AB}|VH\rangle_{EF}|\phi\rangle_{CD}|C_{1}\rangle|C_{3}\rangle|T_{22}T_{32}\rangle)|\alpha_{4}\rangle \nonumber\\
			&&\!\!\!\!\!+[(\beta_{1}\gamma_{1}|HV\rangle_{AB}|HV\rangle_{EF}|\phi\rangle_{CD}|C_{2}\rangle|C_{4}\rangle|T_{11}T_{42}\rangle  \nonumber\\  
			&&\!\!\!\!\!+\beta_{2}\gamma_{1}|VH\rangle_{AB}|HV\rangle_{EF}|\phi\rangle_{CD}|C_{1}\rangle|C_{4}\rangle|T_{21}T_{32}\rangle)|\alpha_{4}e^{2i\theta}\rangle \nonumber\\	
			&&\!\!\!\!\!+(\beta_{1}\gamma_{2}|HV\rangle_{AB}|VH\rangle_{EF}|\phi\rangle_{CD}|C_{2}\rangle|C_{3}\rangle|T_{12}T_{41}\rangle  \nonumber\\
			&&\!\!\!\!\!+\beta_{2}\gamma_{2}|VH\rangle_{AB}|VH\rangle_{EF}|\phi\rangle_{CD}|C_{1}\rangle|C_{3}\rangle|T_{22}T_{31}\rangle)]|\alpha_{4}e^{-2i\theta}\rangle \nonumber\\	
			&&\!\!\!\!\!+[(\beta_{1}\gamma_{1}|HV\rangle_{AB}|HV\rangle_{EF}|\phi\rangle_{CD}|C_{2}\rangle|C_{4}\rangle|T_{12}T_{41}\rangle  \nonumber\\  
			&&\!\!\!\!\!+\beta_{2}\gamma_{1}|VH\rangle_{AB}|HV\rangle_{EF}|\phi\rangle_{CD}|C_{1}\rangle|C_{4}\rangle|T_{22}T_{31}\rangle)|\alpha_{4}e^{3i\theta}\rangle \nonumber\\	
			&&\!\!\!\!\!+(\beta_{1}\gamma_{2}|HV\rangle_{AB}|VH\rangle_{EF}|\phi\rangle_{CD}|C_{2}\rangle|C_{3}\rangle|T_{11}T_{42}\rangle  \nonumber\\
			&&\!\!\!\!\!+\beta_{2}\gamma_{2}|VH\rangle_{AB}|VH\rangle_{EF}|\phi\rangle_{CD}|C_{1}\rangle|C_{3}\rangle|T_{21}T_{32}\rangle)]|\alpha_{4}e^{-3i\theta}\rangle \nonumber\\	
			&&\!\!\!\!\!+[(\beta_{1}\gamma_{1}|HV\rangle_{AB}|HV\rangle_{EF}|\phi\rangle_{CD}|C_{2}\rangle|C_{4}\rangle|T_{12}T_{42}\rangle  \nonumber\\  
			&&\!\!\!\!\!+\beta_{2}\gamma_{1}|VH\rangle_{AB}|HV\rangle_{EF}|\phi\rangle_{CD}|C_{1}\rangle|C_{4}\rangle|T_{22}T_{32}\rangle)|\alpha_{4}e^{5i\theta}\rangle \nonumber\\	
			&&\!\!\!\!\!+(\beta_{1}\gamma_{2}|HV\rangle_{AB}|VH\rangle_{EF}|\phi\rangle_{CD}|C_{2}\rangle|C_{3}\rangle|T_{11}T_{41}\rangle  \nonumber\\
			&&\!\!\!\!\!+\beta_{2}\gamma_{2}|VH\rangle_{AB}|VH\rangle_{EF}|\phi\rangle_{CD}|C_{1}\rangle|C_{3}\rangle|T_{21}T_{31}\rangle)]|\alpha_{4}e^{-5i\theta}\rangle. 
		\end{eqnarray}
	\end{widetext}
	After performing the X-homodyne measurement on the coherent state $\alpha_{4}$, we once again obtain four sets of measurement outcomes, and corresponding feed-forward operations shown in Table \ref{Table:3} are executed based on the measurement outcomes, resulting in the quantum state $|\Phi\rangle_{2}$ collapses into
	\begin{eqnarray}
		\!\!|\Phi\rangle_{2}'&\!=\!& \beta_{1}\gamma_{1}|HV\rangle_{AB}|HV\rangle_{EF}|\phi\rangle_{CD}|C_{2}\rangle|C_{4}\rangle|T_{11}T_{41}\rangle  \nonumber\\  
		&&\!\!\!\!+\beta_{2}\gamma_{1}|VH\rangle_{AB}|HV\rangle_{EF}|\phi\rangle_{CD}|C_{1}\rangle|C_{4}\rangle|T_{21}T_{31}\rangle \nonumber\\	
		&&\!\!\!\!+\beta_{1}\gamma_{2}|HV\rangle_{AB}|VH\rangle_{EF}|\phi\rangle_{CD}|C_{2}\rangle|C_{3}\rangle|T_{12}T_{42}\rangle  \nonumber\\
		&&\!\!\!\!+\beta_{2}\gamma_{2}|VH\rangle_{AB}|VH\rangle_{EF}|\phi\rangle_{CD}|C_{1}\rangle|C_{3}\rangle|T_{22}T_{32}\rangle. \nonumber\\
	\end{eqnarray}

	\begin{table}[htbp]
		\caption{\label{Table:3} The measurement results of the coherent state $|\alpha_{4}\rangle$, and the corresponding single-qubit feed-forward operations.}
		\begin{tabular}{ccr} \hline
			\multicolumn{1}{c}{\begin{tabular}[c]{@{}c@{}}Measurement\\ results\end{tabular}}  &\multicolumn{1}{c}{\begin{tabular}[c]{@{}c@{}}Phase \\modulation\end{tabular}}  & \multicolumn{1}{c}{\begin{tabular}[c]{@{}c@{}}Single-qubit\\ operations\end{tabular}}\\
			\hline
			\multicolumn{1}{c}{0}           &0   & \multicolumn{1}{c}{\begin{tabular}[c]{@{}c@{}}$ I_{T_{11},T_{12}}\otimes I_{T_{21},T_{22}}$\\ $ \otimes I_{T_{31},T_{32}}\otimes I_{T_{41},T_{42}}$\end{tabular}}\\
			\multicolumn{1}{c}{$\pm2\theta$}   &\multicolumn{1}{c}{$\phi(x, 2\theta)_{C_{3}}$}   & \multicolumn{1}{c}{\begin{tabular}[c]{@{}c@{}}$ I_{T_{11},T_{12}}\otimes I_{T_{21},T_{22}}$\\ $ \otimes S_{T_{31},T_{32}}\otimes S_{T_{41},T_{42}}$\end{tabular}}\\
			\multicolumn{1}{c}{$\pm3\theta$}   &\multicolumn{1}{c}{$\phi(x, 3\theta)_{C_{3}}$}   & \multicolumn{1}{c}{\begin{tabular}[c]{@{}c@{}}$ S_{T_{11},T_{12}}\otimes S_{T_{21},T_{22}}$\\ $ \otimes I_{T_{31},T_{32}}\otimes I_{T_{41},T_{42}}$\end{tabular}}\\
			\multicolumn{1}{c}{$\pm5\theta$}   &\multicolumn{1}{c}{$\phi(x, 5\theta)_{C_{3}}$}   & \multicolumn{1}{c}{\begin{tabular}[c]{@{}c@{}}$ S_{T_{11},T_{12}}\otimes S_{T_{21},T_{22}}$\\ $ \otimes S_{T_{31},T_{32}}\otimes S_{T_{41},T_{42}}$\end{tabular}}\\  \hline
		\end{tabular}
	\end{table}

	Thirdly, the target photons CD pass through $\rm X_{1}$ and $\rm X_{2}$, respectively. The evolution of the system can be illustrated as follows
	\begin{eqnarray}
		\!\!|\Phi\rangle_{3}&\!\!=\!\!& \beta_{1}\gamma_{1}|HV\rangle_{AB}|HV\rangle_{EF}|\phi\rangle_{CD}|C_{2}\rangle|C_{4}\rangle|T_{11}T_{41}\rangle  \nonumber\\  
		&&\!\!\!\!\!+\beta_{2}\gamma_{1}|VH\rangle_{AB}|HV\rangle_{EF}|\phi\rangle_{CD}|C_{1}\rangle|C_{4}\rangle|T_{21}T_{31}\rangle \nonumber\\	
		&&\!\!\!\!\!+\beta_{1}\gamma_{2}|HV\rangle_{AB}|VH\rangle_{EF}|\phi\rangle_{CD}|C_{2}\rangle|C_{3}\rangle|T_{12}T_{42}\rangle  \nonumber\\
		&&\!\!\!\!\!+\beta_{2}\gamma_{2}|VH\rangle_{AB}|VH\rangle_{EF}\overline{|\phi\rangle}_{CD}|C_{1}\rangle|C_{3}\rangle|T_{22}T_{32}\rangle.   \nonumber\\
	\end{eqnarray}

	Fourthly, the target photons CD pass through the two path couplers in Figure \ref{fig1}b, merging the paths ($T_{11}$, $T_{12}$), ($T_{21}$, $T_{22}$), ($T_{31}$, $T_{32}$) and ($T_{41}$, $T_{42}$) into the paths $T_{1}$, $T_{2}$, $T_{3}$ and $T_{4}$, respectively. Then, the quantum state becomes into
	\begin{eqnarray}
		|\Phi\rangle_{4}&\!=\!& \beta_{1}\gamma_{1}|HV\rangle_{AB}|HV\rangle_{EF}|\phi\rangle_{CD}|C_{2}\rangle|C_{4}\rangle|T_{1}T_{4}\rangle  \nonumber\\  
		&&\!\!\!\!+\beta_{2}\gamma_{1}|VH\rangle_{AB}|HV\rangle_{EF}|\phi\rangle_{CD}|C_{1}\rangle|C_{4}\rangle|T_{2}T_{3}\rangle \nonumber\\	
		&&\!\!\!\!+\beta_{1}\gamma_{2}|HV\rangle_{AB}|VH\rangle_{EF}|\phi\rangle_{CD}|C_{2}\rangle|C_{3}\rangle|T_{1}T_{4}\rangle  \nonumber\\
		&&\!\!\!\!+\beta_{2}\gamma_{2}|VH\rangle_{AB}|VH\rangle_{EF}\overline{|\phi\rangle}_{CD}|C_{1}\rangle|C_{3}\rangle|T_{2}T_{3}\rangle.  \nonumber\\
	\end{eqnarray}
	
	\begin{figure*}[ht]
		\centering
		\includegraphics[width=13.5cm,angle=0]{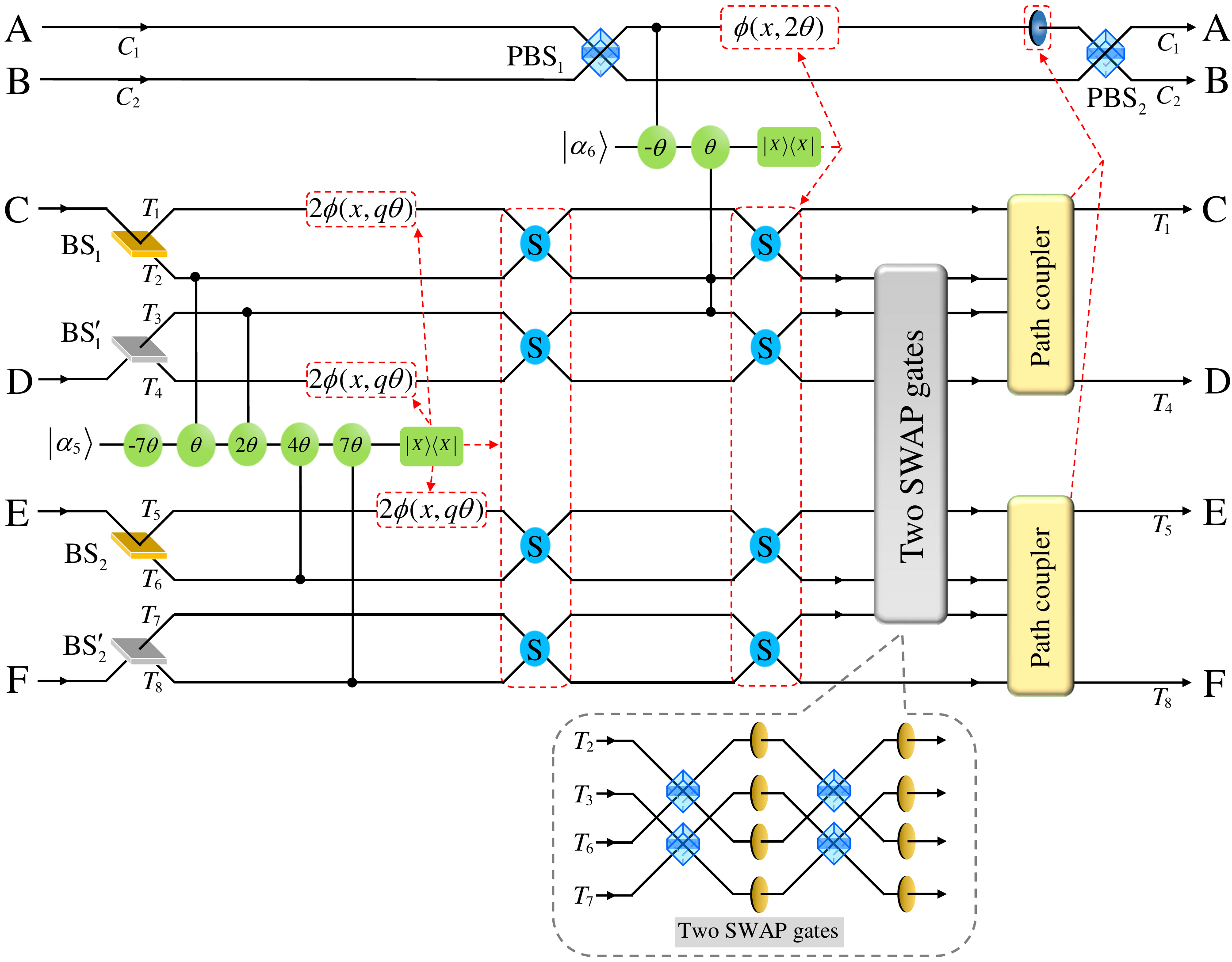}
		\caption{\label{fig4} Schematic diagram of the Fredkin gate in DFS based on cross-Kerr nonlinearities.
			$2\phi(x,q\theta)(q=1, 2, ..., 7)$ represents the phase modulation.}
	\end{figure*}
	
	Finally, the target photons qubits CD pass through the third path coupler again, merging paths $T_{1}$ and $T_{2}$ ($T_{3}$ and $T_{4}$) into paths $T_{1}$ ($T_{4}$), and meanwhile the control photons AB and EF pass through $\rm PBS_{3}$ and $\rm PBS_{4}$, respectively, resulting into
	\begin{eqnarray}  \label{eq17}
		|\Phi\rangle_{5}&\!\!=\!\!& (\beta_{1}\gamma_{1}|HV\rangle_{AB}|HV\rangle_{EF}|\phi\rangle_{CD}  \nonumber\\  
		&&\!\!+\beta_{2}\gamma_{1}|VH\rangle_{AB}|HV\rangle_{EF}|\phi\rangle_{CD} \nonumber\\	
		&&\!\!+\beta_{1}\gamma_{2}|HV\rangle_{AB}|VH\rangle_{EF}|\phi\rangle_{CD}  \nonumber\\
		&&\!\!+\beta_{2}\gamma_{2}|VH\rangle_{AB}|VH\rangle_{EF}\overline{|\phi\rangle}_{CD})\nonumber\\
		&&\!\!\otimes|C_{1}C_{2}\rangle|C_{3}C_{4}\rangle|T_{1}T_{4}\rangle \nonumber\\
		&\!\!=\!\!&(\beta_{1}\gamma_{1}|\overline{0}\rangle_{AB}|\overline{0}\rangle_{EF}|\phi\rangle_{CD} \nonumber\\
		&&\!\!+\beta_{1}\gamma_{2}|\overline{0}\rangle_{AB}|\overline{1}\rangle_{EF}|\phi\rangle_{CD}           \nonumber\\
		&&\!\!+\beta_{2}\gamma_{1}|\overline{1}\rangle_{AB}|\overline{0}\rangle_{EF}|\phi\rangle_{CD}  \nonumber\\
		&&\!\!+\beta_{2}\gamma_{2}|\overline{1}\rangle_{AB}|\overline{1}\rangle_{EF}\overline{|\phi\rangle}_{CD}) \nonumber\\
		&&\!\!\otimes|C_{1}C_{2}\rangle|C_{3}C_{4}\rangle|T_{1}T_{4}\rangle.
	\end{eqnarray}
	Apparently, based on Eq. (\ref{eq17}) the three-logic-qubit Toffoli gate encoding with six photons ABCDEF in DFS is  fulfilled perfectly with the success probability close to 1,
	by performing appropriate feed-forward operations in Tables \ref{Table:1}-\ref{Table:3},
	where the Toffoli gate flips the states of the target photons CD if and only if the control photons AB and EF are both in the state $|\overline{1}\rangle_{AB}|\overline{1}\rangle_{EF}$, and has no change otherwise.

	\subsection{ Fredkin gate}\label{sec:3.3}

	The quantum circuit of the three-logic-qubit Fredkin gate is depicted in Figure \ref{fig4}, where the
	initial states of  three logical qubits are the same as the one of the Toffoli gate, with four photons CD and EF serving as two target logical qubits
	and photons AB as a unique control logical qubit.
	Firstly, four target photons CDEF traverse $\rm BS_{1}$, $\rm BS_{1}'$, $\rm BS_{2}$, and $\rm BS_{2}'$, respectively, and then undergo the cross-Kerr nonlinearity. The evolution of system is illustrated by
	\begin{eqnarray}  \label{eq21}
		|\Psi\rangle_{1}&=&|\phi\rangle_{AB}|\phi\rangle_{CD}|\phi\rangle_{EF}
		|C_{1}C_{2}\rangle|T_{1}T_{4}\rangle|T_{5}T_{8}\rangle \nonumber\\
		&\stackrel{kerr}{\longrightarrow}&
		|\phi\rangle_{AB}\otimes |\phi\rangle_{CD} \otimes|\phi\rangle_{EF}           \nonumber\\
		&&\otimes(|C_{1}C_{2}\rangle|T_{1}T_{4}\rangle|T_{5}T_{8}\rangle|\alpha_{5}\rangle\nonumber\\
		&&
		+|C_{1}C_{2}\rangle|T_{2}T_{3}\rangle|T_{6}T_{7}\rangle|\alpha_{5}\rangle     \nonumber\\
		&&+|C_{1}C_{2}\rangle|T_{2}T_{4}\rangle|T_{5}T_{8}\rangle|\alpha_{5}e^{i\theta}\rangle\nonumber\\
		&&
		+|C_{1}C_{2}\rangle|T_{1}T_{3}\rangle|T_{6}T_{7}\rangle|\alpha_{5}e^{-i\theta}\rangle     \nonumber\\
		&&+|C_{1}C_{2}\rangle|T_{1}T_{3}\rangle|T_{5}T_{8}\rangle|\alpha_{5}e^{2i\theta}\rangle\nonumber\\
		&&
		+|C_{1}C_{2}\rangle|T_{2}T_{4}\rangle|T_{6}T_{7}\rangle|\alpha_{5}e^{-2i\theta}\rangle     \nonumber\\
		&&+|C_{1}C_{2}\rangle|T_{2}T_{3}\rangle|T_{5}T_{8}\rangle|\alpha_{5}e^{3i\theta}\rangle\nonumber\\
		&&
		+|C_{1}C_{2}\rangle|T_{1}T_{4}\rangle|T_{6}T_{7}\rangle|\alpha_{5}e^{-3i\theta}\rangle     \nonumber\\
		&&+|C_{1}C_{2}\rangle|T_{1}T_{4}\rangle|T_{6}T_{8}\rangle|\alpha_{5}e^{4i\theta}\rangle\nonumber\\
		&&
		+|C_{1}C_{2}\rangle|T_{2}T_{3}\rangle|T_{5}T_{7}\rangle|\alpha_{5}e^{-4i\theta}\rangle     \nonumber\\
		&&+|C_{1}C_{2}\rangle|T_{2}T_{4}\rangle|T_{6}T_{8}\rangle|\alpha_{5}e^{5i\theta}\rangle\nonumber\\
		&&
		+|C_{1}C_{2}\rangle|T_{1}T_{3}\rangle|T_{5}T_{7}\rangle|\alpha_{5}e^{-5i\theta}\rangle     \nonumber\\
		&&+|C_{1}C_{2}\rangle|T_{1}T_{3}\rangle|T_{6}T_{8}\rangle|\alpha_{5}e^{6i\theta}\rangle\nonumber\\
		&&
		+|C_{1}C_{2}\rangle|T_{2}T_{4}\rangle|T_{5}T_{7}\rangle|\alpha_{5}e^{-6i\theta}\rangle     \nonumber\\
		&&+|C_{1}C_{2}\rangle|T_{2}T_{3}\rangle|T_{6}T_{8}\rangle|\alpha_{5}e^{7i\theta}\rangle\nonumber\\
		&&
		+|C_{1}C_{2}\rangle|T_{1}T_{4}\rangle|T_{5}T_{7}\rangle|\alpha_{5}e^{-7i\theta}\rangle).
	\end{eqnarray}
	If the outcome is zero phase shift, the state of the composite system can be denoted as
	\begin{eqnarray}   \label{eq22}
		\!\!\!\!\!\!|\Psi\rangle_{1}'\!&=&\!
		|\phi\rangle_{AB}|\phi\rangle_{CD}|\phi\rangle_{EF}(|C_{1}C_{2}\rangle|T_{1}T_{4}\rangle|T_{5}T_{8}\rangle
		\nonumber\\
		&&\!\!\!\!+|C_{1}C_{2}\rangle|T_{2}T_{3}\rangle|T_{6}T_{7}\rangle).
	\end{eqnarray}
	Otherwise, nonzero phase shifts $(\pm\theta, \pm2\theta, \cdots, \pm7\theta)$ are obtained. The
	phase shift $2\phi(x, q\theta)$ $(q= 1,\cdots, 7)$ operation should be performed, along with the corresponding feed-forward single-qubit operations as shown in Table \ref{Table:4} to achieve the desired state in Eq. (\ref{eq22}).

	\begin{table}[htbp]
		\caption{\label{Table:4} The measurement results of the coherent state $|\alpha_{5}\rangle$, and the corresponding single-qubit feed-forward operations.}
		\begin{tabular}{ccr} \hline
			\multicolumn{1}{c}{\begin{tabular}[c]{@{}c@{}}Measurement\\ results\end{tabular}}  &\multicolumn{1}{c}{\begin{tabular}[c]{@{}c@{}}Phase \\modulation\end{tabular}}  & \multicolumn{1}{c}{\begin{tabular}[c]{@{}c@{}}Single-qubit\\ operations\end{tabular}}\\
			\hline
			\multicolumn{1}{c}{0}           &0  & \multicolumn{1}{c}{\begin{tabular}[c]{@{}c@{}}$ I_{T_{1},T_{2}}$\\ $ \otimes I_{T_{3},T_{4}}\otimes I_{T_{5},T_{6}}$\end{tabular}}\\
			\multicolumn{1}{c}{$\pm\theta$}   &\multicolumn{1}{c}{$2\phi(x, \theta)_{T_{1}}$}  & \multicolumn{1}{c}{\begin{tabular}[c]{@{}c@{}}$ S_{T_{1},T_{2}}$\\ $ \otimes I_{T_{3},T_{4}}\otimes I_{T_{5},T_{6}}$\end{tabular}}\\
			\multicolumn{1}{c}{$\pm2\theta$}   &\multicolumn{1}{c}{$2\phi(x, 2\theta)_{T_{4}}$}  & \multicolumn{1}{c}{\begin{tabular}[c]{@{}c@{}}$ I_{T_{1},T_{2}}$\\ $ \otimes S_{T_{3},T_{4}}\otimes I_{T_{5},T_{6}}$\end{tabular}}\\
			\multicolumn{1}{c}{$\pm3\theta$}   &\multicolumn{1}{c}{$2\phi(x, 3\theta)_{T_{4}}$}  & \multicolumn{1}{c}{\begin{tabular}[c]{@{}c@{}}$ I_{T_{1},T_{2}}$\\ $ \otimes S_{T_{3},T_{4}}\otimes S_{T_{5},T_{6}}$\end{tabular}}\\
			\multicolumn{1}{c}{$\pm4\theta$}   &\multicolumn{1}{c}{$2\phi(x, 4\theta)_{T_{5}}$}  & \multicolumn{1}{c}{\begin{tabular}[c]{@{}c@{}}$ I_{T_{1},T_{2}}$\\ $ \otimes I_{T_{3},T_{4}}\otimes S_{T_{5},T_{6}}$\end{tabular}}\\
			\multicolumn{1}{c}{$\pm5\theta$}   &\multicolumn{1}{c}{$2\phi(x, 5\theta)_{T_{5}}$}  & \multicolumn{1}{c}{\begin{tabular}[c]{@{}c@{}}$ S_{T_{1},T_{2}}$\\ $ \otimes I_{T_{3},T_{4}}\otimes S_{T_{5},T_{6}}$\end{tabular}}\\
			\multicolumn{1}{c}{$\pm6\theta$}   &\multicolumn{1}{c}{$2\phi(x, 6\theta)_{T_{5}}$}  & \multicolumn{1}{c}{\begin{tabular}[c]{@{}c@{}}$ I_{T_{1},T_{2}}$\\ $ \otimes S_{T_{3},T_{4}}\otimes S_{T_{5},T_{6}}$\end{tabular}}\\
			\multicolumn{1}{c}{$\pm7\theta$}   &\multicolumn{1}{c}{$2\phi(x, 7\theta)_{T_{5}}$}  & \multicolumn{1}{c}{\begin{tabular}[c]{@{}c@{}}$ S_{T_{1},T_{2}}$\\ $ \otimes S_{T_{3},T_{4}}\otimes S_{T_{5},T_{6}}$\end{tabular}}\\ \hline
		\end{tabular}
	\end{table}

	Secondly, the control photons AB traverse $\rm PBS_{1}$, and the evolution process of they interacting with the coherent state $|\alpha_{6}\rangle$ is expressed as
	\begin{eqnarray}
		|\Psi\rangle_{2}\!&=&\!
		(\beta_{1}|HV\rangle_{AB}|\phi\rangle_{CD}|\phi\rangle_{EF} \nonumber\\
		&&\!\!\!\!\otimes|C_{2}\rangle|T_{1}T_{4}\rangle|T_{5}T_{8}\rangle    \nonumber\\
		&&\!\!\!\!+\beta_{2}|VH\rangle_{AB}|\phi\rangle_{CD}|\phi\rangle_{EF} \nonumber\\
		&&\!\!\!\!\otimes
		|C_{1}\rangle|T_{2}T_{3}\rangle|T_{6}T_{7}\rangle)|\alpha_{6}\rangle    \nonumber\\
		&&\!\!\!\!+\beta_{1}|HV\rangle_{AB}|\phi\rangle_{CD}|\phi\rangle_{EF} \nonumber\\
		&&\!\!\!\!\otimes
		|C_{2}\rangle|T_{2}T_{3}\rangle|T_{6}T_{7}\rangle|\alpha_{6}e^{2i\theta}\rangle    \nonumber\\
		&&\!\!\!\!+\beta_{2}|VH\rangle_{AB}|\phi\rangle_{CD}|\phi\rangle_{EF} \nonumber\\
		&&\!\!\!\!\otimes
		|C_{1}\rangle|T_{1}T_{4}\rangle|T_{5}T_{8}\rangle|\alpha_{6}e^{-2i\theta}\rangle.   
	\end{eqnarray}
	If no phase shift is induced in the coherent state $|\alpha_{6}\rangle$, there is nothing to do. However, if a phase shift of $\pm2\theta$ is induced, a feed-forward $\phi(x, 2\theta)$ operation is applied on path $T_{1}$, followed by four SWAP gates applied on some paths $(T_{1},T_{2})$, $(T_{3},T_{4})$, $(T_{5},T_{6})$ and $(T_{7},T_{8})$. Then, the finally state is given by
	\begin{eqnarray}
		|\Psi\rangle_{2}'\!&=&\!
		\beta_{1}|HV\rangle_{AB}|\phi\rangle_{CD}|\phi\rangle_{EF}
		|C_{2}\rangle|T_{1}T_{4}\rangle|T_{5}T_{8}\rangle  \nonumber\\
		&&\!\!+\beta_{2}|VH\rangle_{AB}|\phi\rangle_{CD}|\phi\rangle_{EF}
		|C_{1}\rangle|T_{2}T_{3}\rangle|T_{6}T_{7}\rangle.\nonumber\\
	\end{eqnarray}

	Thirdly, two SWAP gates are executed on the target photons CE on two paths $|T_{2}T_{6}\rangle$ and the target photons DF on two paths $|T_{3}T_{7}\rangle$, respectively, and the quantum state is evolved into
	\begin{eqnarray}
		\!\!\!|\Psi\rangle_{3}\!&=&\!
		\beta_{1}|\overline{0}\rangle_{AB}|\phi\rangle_{CD}|\phi\rangle_{EF}
		|C_{2}\rangle|T_{1}T_{4}\rangle|T_{5}T_{8}\rangle  \nonumber\\
		&&\!+\beta_{2}|\overline{1}\rangle_{AB}\overline{|\phi\rangle_{CD}|\phi\rangle_{EF}}
		|C_{1}\rangle|T_{2}T_{3}\rangle|T_{6}T_{7}\rangle,\nonumber\\
	\end{eqnarray}
	where
	\begin{eqnarray}
		\overline{|\phi\rangle_{CD}|\phi\rangle_{EF}}&=&(\gamma_{1}|\overline{0}\rangle+\gamma_{2}|\overline{1}\rangle)_{CD}\nonumber\\
		&&\!\!\!\otimes
		(\delta_{1}|\overline{0}\rangle+\delta_{2}|\overline{1}\rangle)_{EF}.
	\end{eqnarray}

	Finally, the control photons AB traverse $\rm PBS_{2}$ and the target photons CD and EF traverse two path couplers in Figure \ref{fig1}b to simplify the circuits, respectively, and the desirous state is obtained
	\begin{eqnarray}  \label{eq24}
		|\Psi\rangle_{4}\!\!\!&=&\!\!\!
		(\beta_{1}|\overline{0}\rangle_{AB}|\phi\rangle_{CD}|\phi\rangle_{EF} +\beta_{2}|\overline{1}\rangle_{AB}\overline{|\phi\rangle_{CD}|\phi\rangle_{EF}})  \nonumber\\
		&&\!\!\!\otimes|C_{1}C_{2}\rangle|T_{1}T_{4}\rangle|T_{5}T_{8}\rangle.
	\end{eqnarray}
	Apparently, based on Eq. (\ref{eq24}), the three-logic-qubit Fredkin gate encoding with six photons ABCDEF in DFS is  fulfilled perfectly with the success probability close to 1,
	by performing appropriate feed-forward operations in Tables \ref{Table:1} and \ref{Table:4},
	where the Fredkin gate swaps the states of the target photons CD and EF if and only if the control photons AB is in the state $|\overline{1}\rangle_{AB}$, and has no change otherwise.

	\section{\label{sec:4} Success probabilities and fidelities of quantum gates regard to photon loss}
	
	\begin{figure}[htbp]
		\includegraphics[width=8.2cm,angle=0]{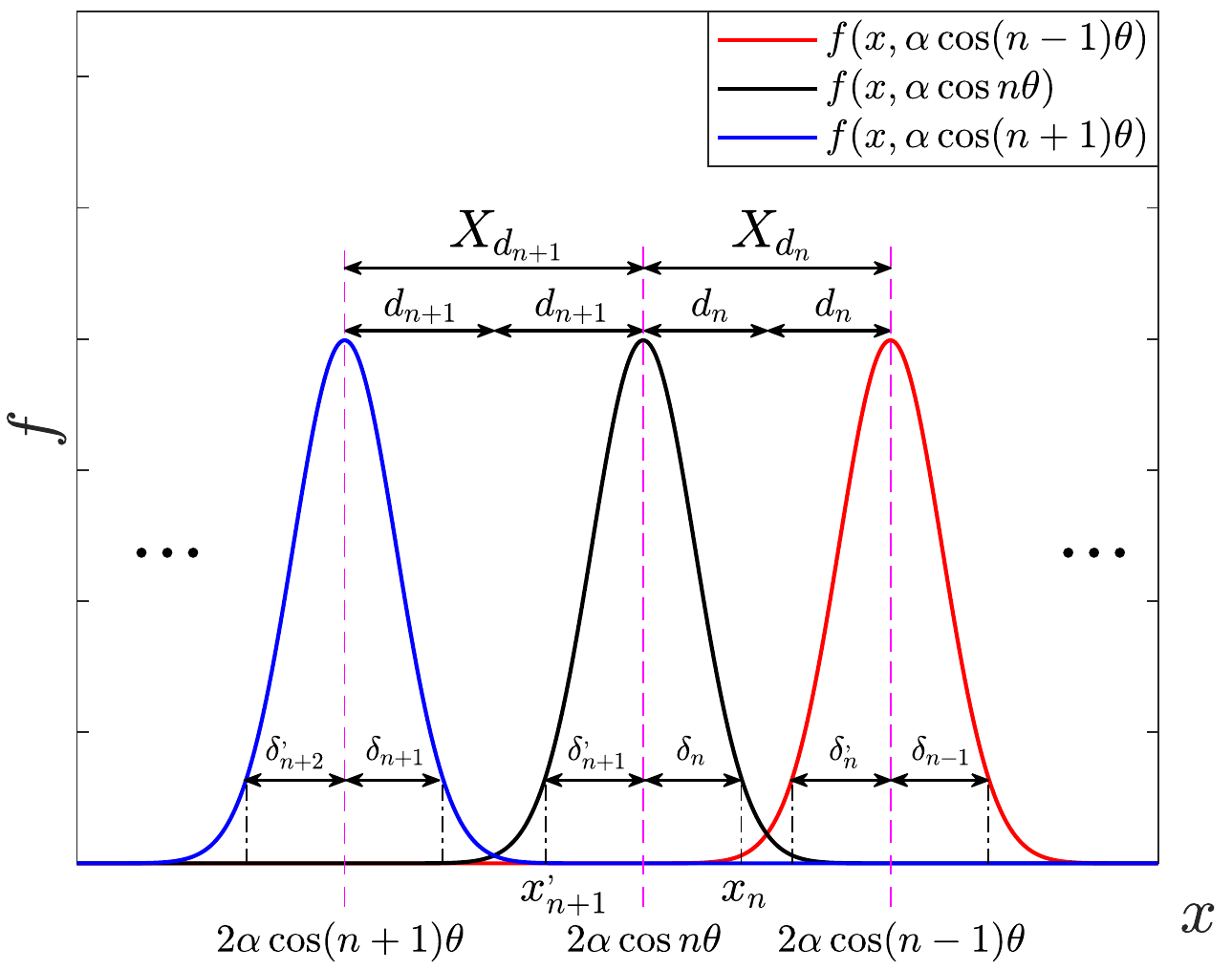}
		\caption{\label{fig5} Gaussian function curves $f(x, \alpha\cos(n-1)\theta)$ (red line), $f(x, \alpha\cos n\theta)$ (black line), and $f(x, \alpha\cos(n+1)\theta)$ (blue line). 		
			The symbol $X_{d_{n}}$ stands for the distance between the peaks of Gaussian function curves $f(x, \alpha\cos(n-1)\theta)$ and $f(x, \alpha\cos n\theta)$. The symbol $X_{d_{n+1}}$ represents the distance between the peaks of Gaussian function curves $f(x, \alpha\cos n\theta)$ and $f(x, \alpha\cos (n+1)\theta)$.
			$X_{d_{n}}=2d_{n}$ and $X_{d_{n+1}}=2d_{n+1}$.		
			$x_{n}$ and $x'_{n+1}$ are measurement outcomes of X Homodyne measurement. $\delta_{n}$ ($\delta'_{n+1}$) is the distance between measurement outcomes $x_{n}$ ($x'_{n+1}$) and the peak point ($x=2\alpha\cos n\theta$) of the Gaussian function curve $f(x,\alpha\cos n\theta)$.}
	\end{figure}
	
	\begin{figure*}[htbp]
		\centering
		\subfigure{
			\includegraphics[width=6.2cm]{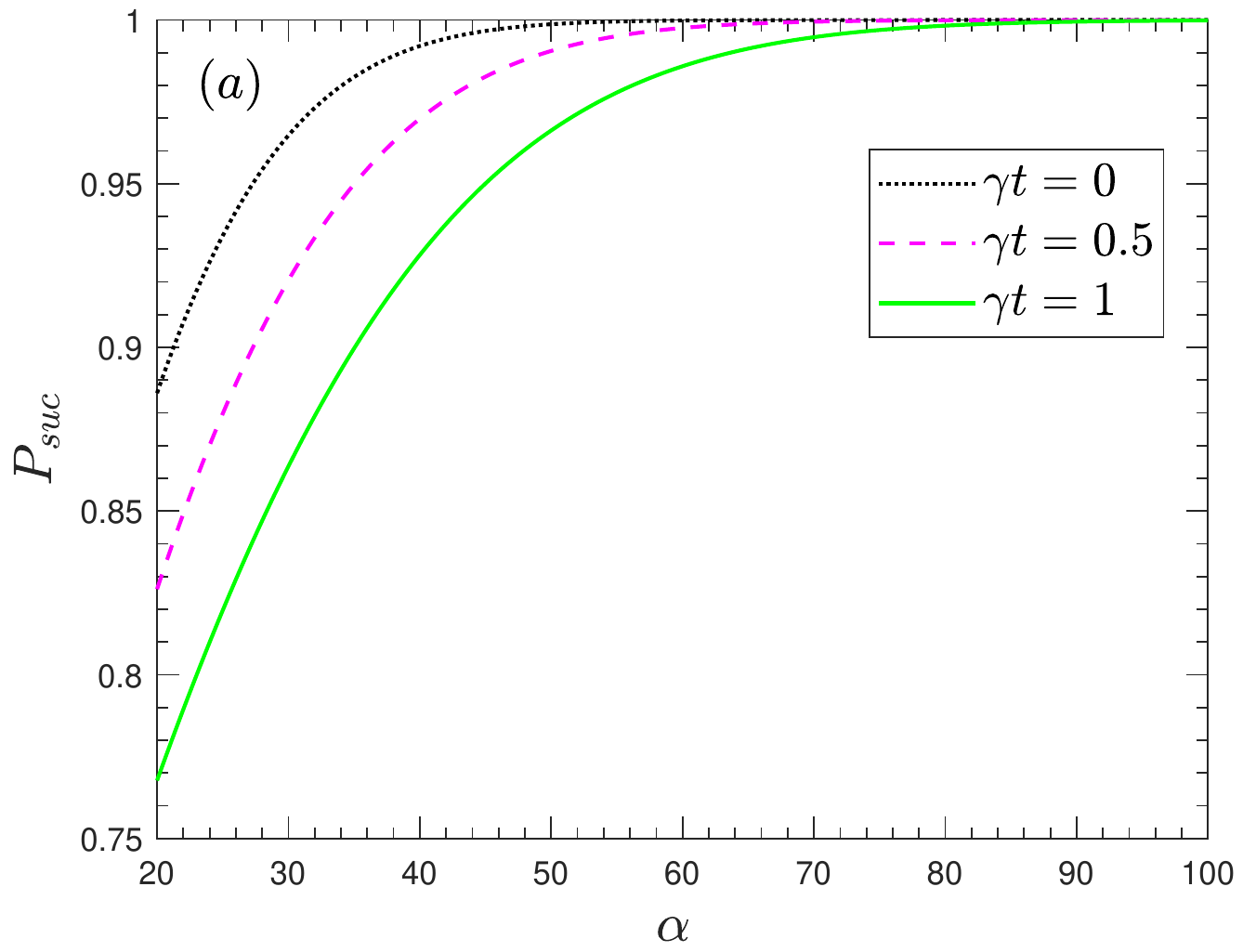}
		}
		\subfigure{
			\includegraphics[width=6.2cm]{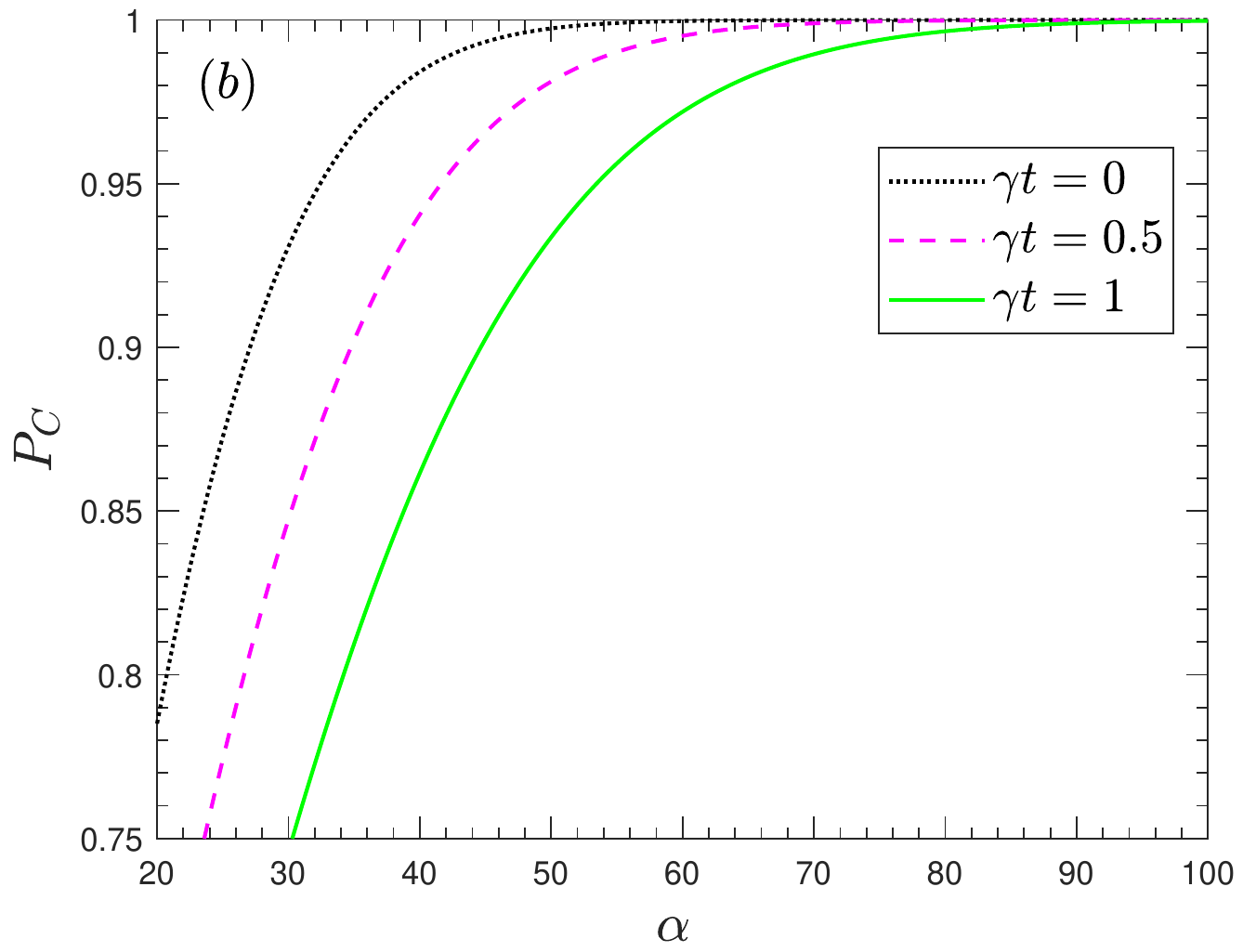}
		}
		\subfigure{
			\includegraphics[width=6.2cm]{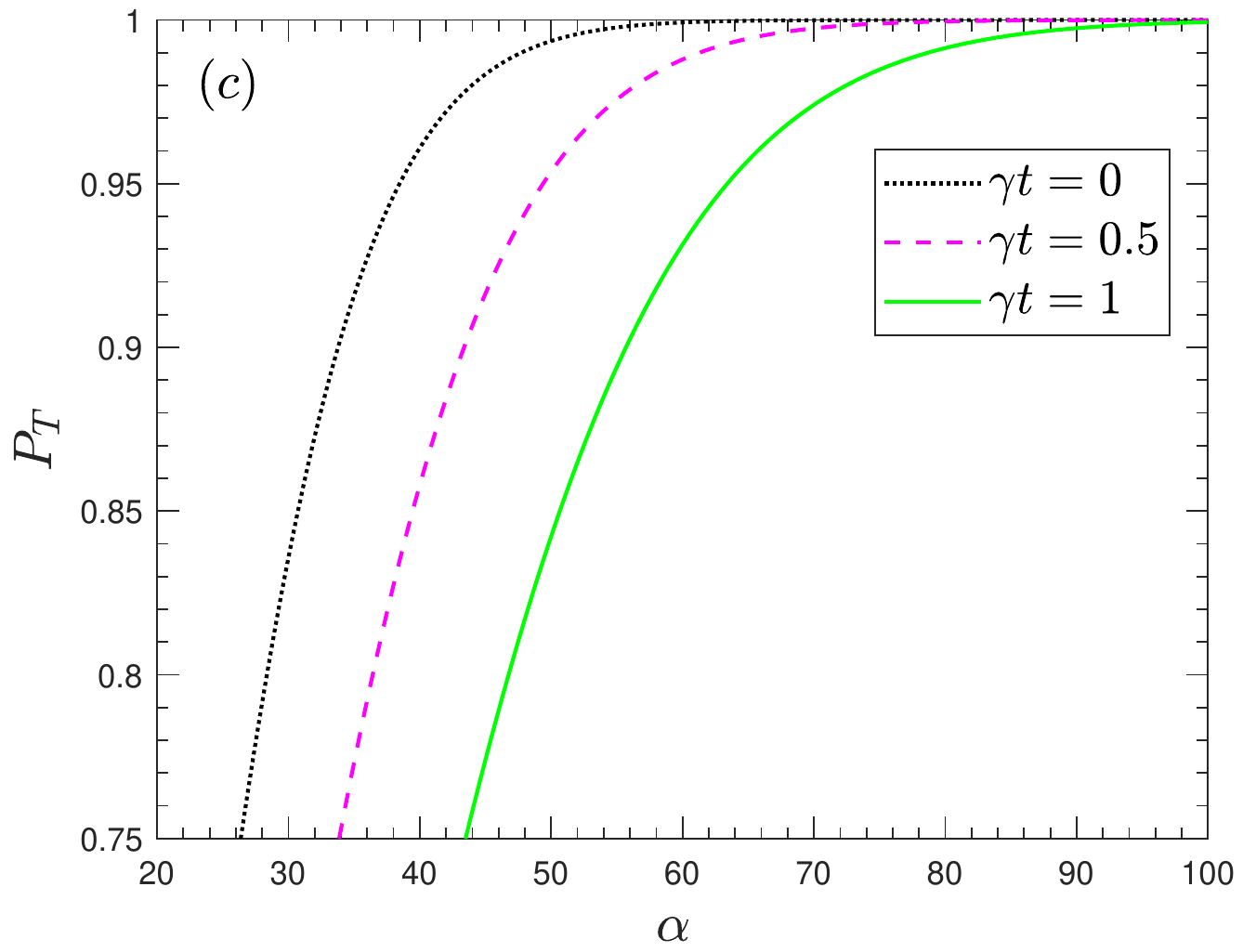}
		}
		\subfigure{
			\includegraphics[width=6.2cm]{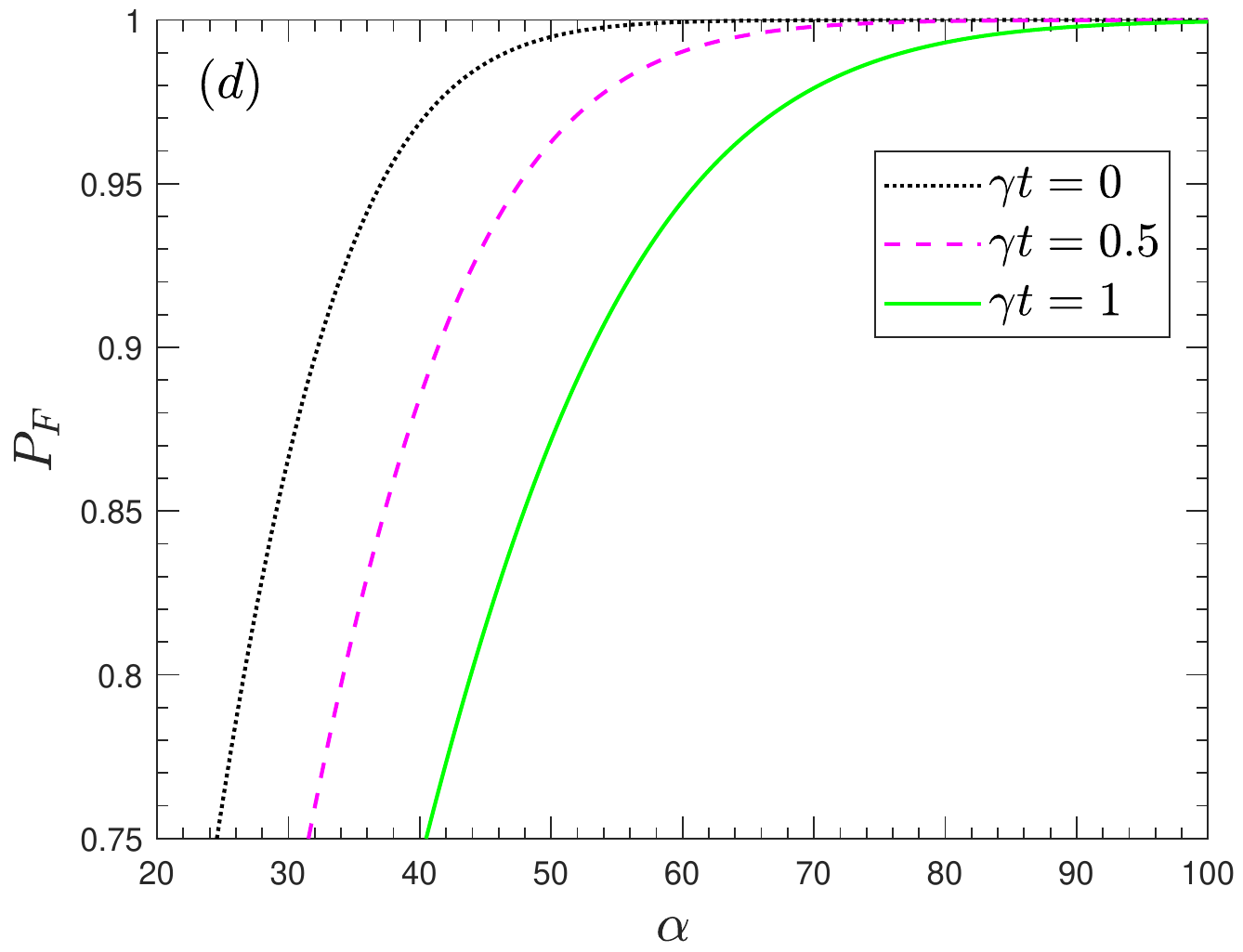}
		}
		\caption{ 			
			(a) The success probability  $P_{suc} (\alpha,\gamma t)$ of the X-homodyne measurement of the path coupler vs the amplitude $\alpha$
			for the fixed dimensionless parameters
			$\theta=0.35$ \cite{0.35} with the various conditions  $\gamma t=0$ (black dotted line), $\gamma t=0.5$ (purple dashed line), and $\gamma t=1$ (green line), respectively.
			(b)-(d) show the corresponding success probabilities of our logic gates with the same condition as (a).
		}
		\label{fig6}
	\end{figure*}

	In view of cross-Kerr nonlinearities, the principles of three schemes for CNOT, Toffoli, and Fredkin gates in DFS to overcome decoherence,
	are discussed in detail in Sec. \ref{sec:3}. Besides, the path couplers of three quantum logical gates are employed to simplify the complexity of quantum circuits and halve the number of the paths of the photons without destroying the polarization information of these photons.
	As the function conforming to $f(x, \alpha \cos\theta)=
	(2\pi)^{-1/4}{\rm exp}[-(x-2\alpha\cos\theta)^{2}/4]$ cannot differentiate phase shift $\pm\theta$ of the coherent states, the X-homodyne measurement can obtain four scenarios of phase shifts, i.e., $0, \pm2\theta, \pm3\theta, \pm5\theta$ from the corresponding coherent states $|\alpha_{1}\rangle, |\alpha_{2}\rangle, |\alpha_{3}\rangle, $ and $|\alpha_{4}\rangle$ for implementation of the path couplers, two-logic-qubit CNOT gate, and three-logic-qubit Toffoli gate.
	It can also distinguish eight scenarios of phase shifts, i.e., $0, \pm\theta, \pm2\theta, \cdots, \pm7\theta$ from the coherent states $|\alpha_{5}\rangle$ and two scenarios of phase shifts, i.e., $0, \pm2\theta$ from the coherent state $|\alpha_{6}\rangle$ for implementation of the three-logic-qubit Fredkin gate.
	Next, we analyze the success probabilities and fidelities of our quantum gates.

	\begin{figure*}[htbp]
		\centering
		\subfigure{
			\includegraphics[width=6.2cm]{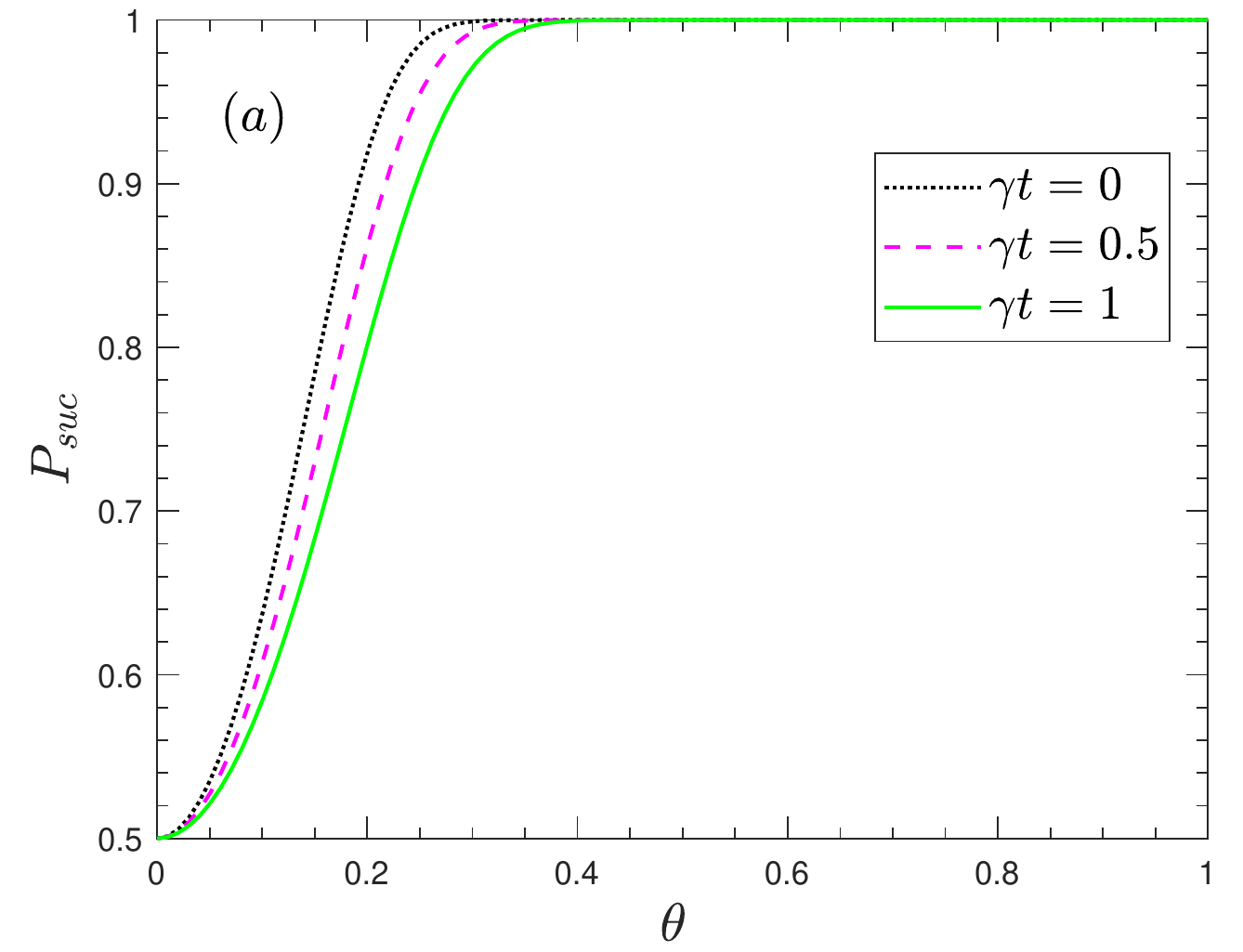}
		}
		\subfigure{
			\includegraphics[width=6.2cm]{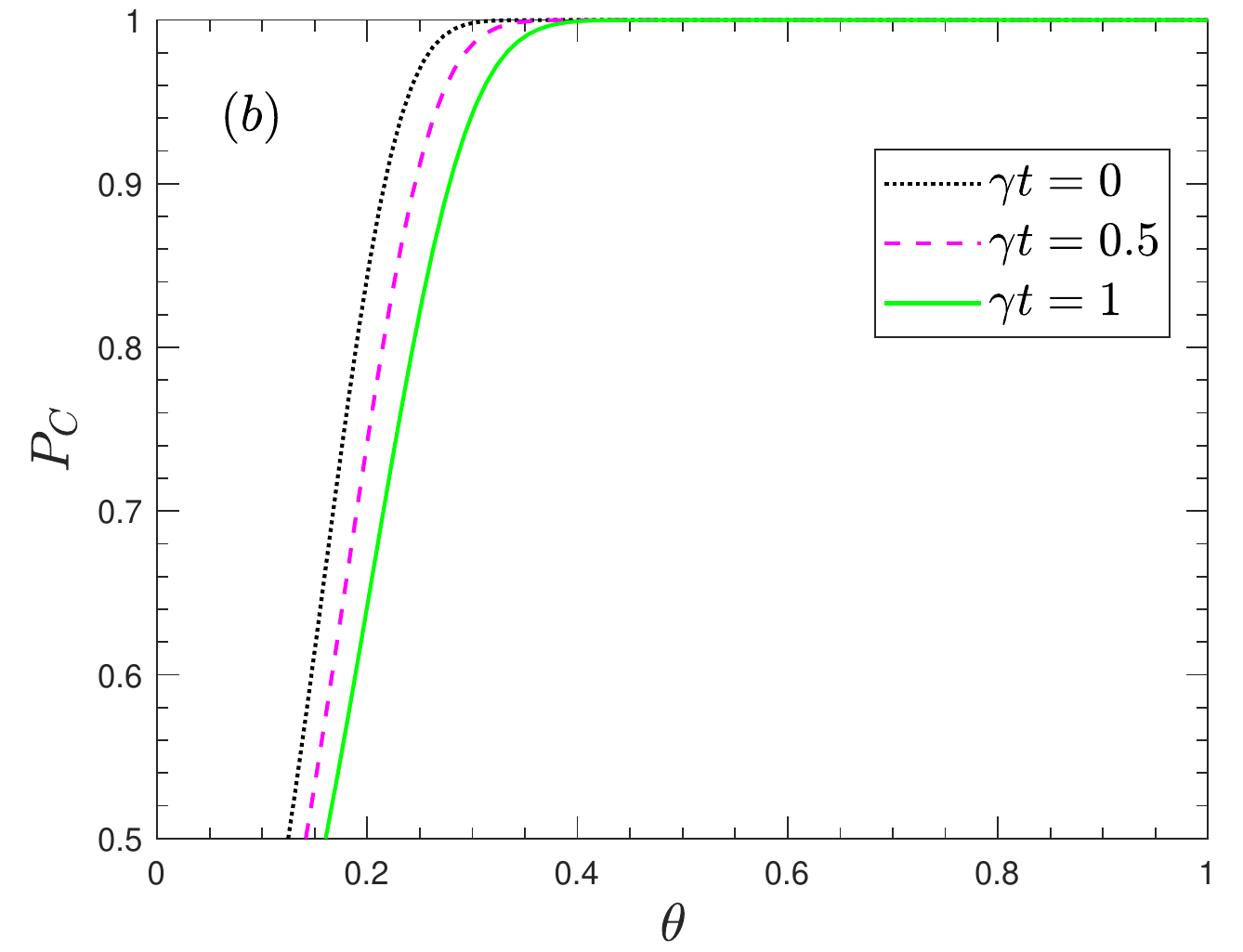}
		}
		\subfigure{
			\includegraphics[width=6.2cm]{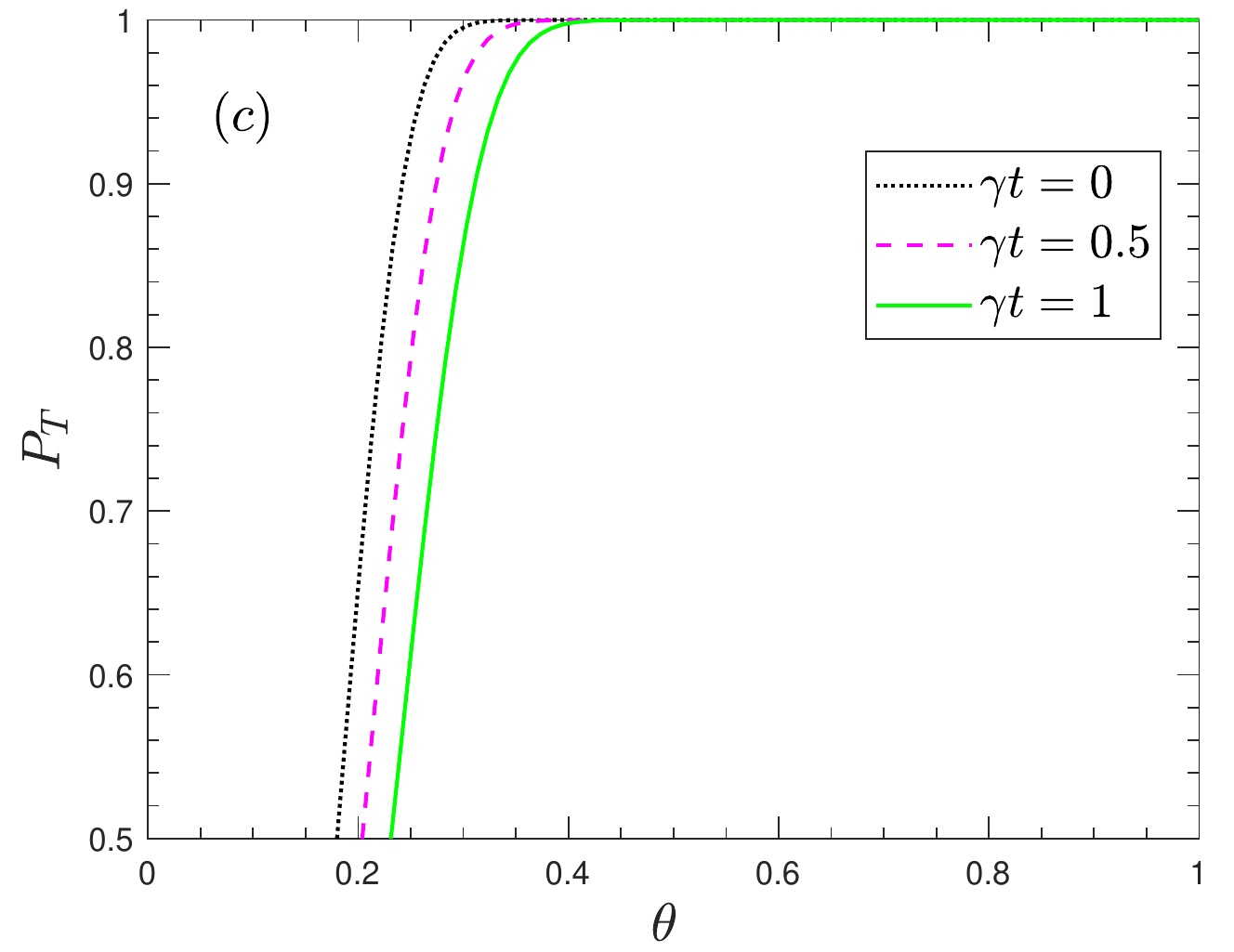}
		}
		\subfigure{
			\includegraphics[width=6.2cm]{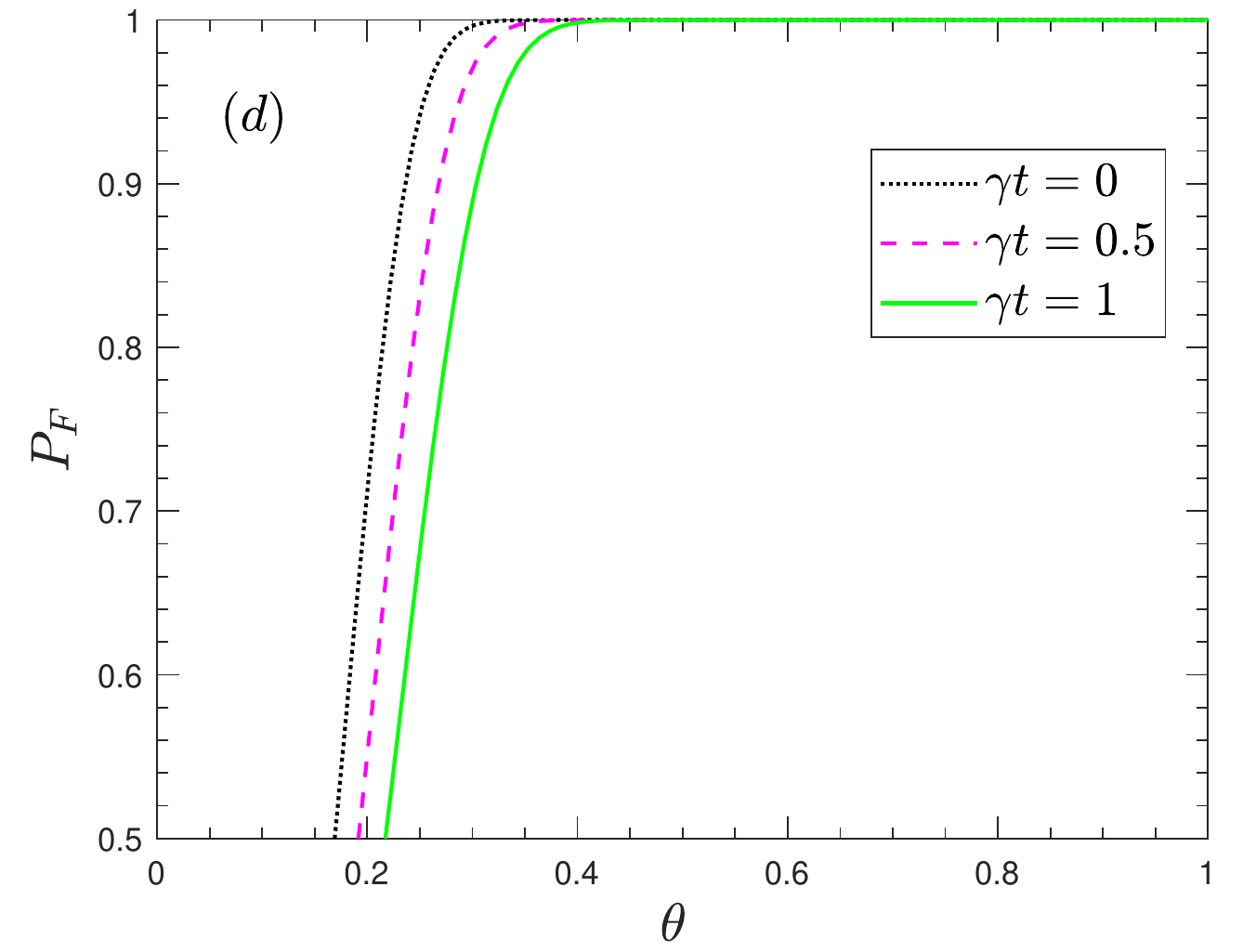}
		}
		\caption{ 			
			(a)The success probability  $P_{suc} (\theta,\gamma t)$ of the X-homodyne measurement of the path coupler vs  the phase shift $\theta$ for the fixed amplitude $\alpha=70$ with the various conditions  $\gamma t=0$ (black dotted line), $\gamma t=0.5$ (purple dashed line), and $\gamma t=1$ (green line), respectively.
			(b)-(d) show the corresponding success probabilities of our logic gates with the same condition as (a).
		}
		\label{fig7}
	\end{figure*}

	\subsection{Success probabilities of quantum gates regard to photon loss}
	The success probabilities $P_{suc}$ of the our protocols relies on the precise measurement of the X-quadrature homodyne.
	The Gaussian function curves $f(x, \alpha\cos(n-1)\theta)$ and $f(x, \alpha\cos n\theta)$ partially overlap, resulting in errors between phase shifts $(n-1)\theta$ and $n\theta$ shown in Figure \ref{fig5}. The symbol $X_{d_{n}}$ stands for the distance between the peaks of Gaussian function curves $f(x, \alpha\cos(n-1)\theta)$ and $f(x, \alpha\cos n\theta)$. Besides, the symbol $X_{d_{n+1}}$ represents the distance between the peaks of Gaussian function curves $f(x, \alpha\cos(n)\theta)$ and $f(x, \alpha\cos (n+1)\theta)$. The photon loss of the probe coherent field in the nonlinear Kerr medium leads to $\alpha\rightarrow A\alpha$, where
	$\gamma$ is the decay constant and $A=e^{-\frac{1}{2}\gamma t}$ is the coherence parameter reducing the amplitude of coherent state and making the original
	pure state to evolve into a mixed state of the photons.
	Considering the photon dissipation of the coherent state, the success probability $P_{suc}$ of the path coupler with once interaction can be calculated by  \cite{Tgate-Dong}
	\begin{eqnarray}  \label{eq31}
		P_{suc}(\alpha,\theta,\gamma t)=1-\frac{1}{2}\rm erfc[\frac{\emph A\alpha(1-\cos\theta)}{\sqrt{2}}],
	\end{eqnarray}
	where erfc(x) is a Gauss complementary error function.
	The factors that determine the success probability $P_{suc}$ include $\alpha$, $\theta$, and $\gamma t$.
	Considering the cascade effects of the homodyne detection, the
	application of twice, five times, and four times X-homodyne measurements to realize
	the CNOT gate, Toffoli gate, and Fredkin gate, respectively, so
	the success probabilities of the CNOT gate ($P_{C}$), Toffoli gate ($P_{T}$), and Fredkin gate ($P_{F}$) can be calculated as $P^{2}_{suc}$, $P^{5}_{suc}$, and $P^{4}_{suc}$ ignoring the imperfect influence of linear optical elements.
	Firstly, we discuss the success probability  $P_{suc} (\alpha,\gamma t)$ of the precise measurement of the X-homodyne detectors vs amplitude $\alpha$ with different conditions $\gamma t=0$, $\gamma t=0.5$, and $\gamma t=1$ for fixed the
	parameters $\theta=0.35$ \cite{0.35}, depicted in Figure \ref{fig6}a.  The success probability $P_{C}$, $P_{T}$, and $P_{F}$ of the CNOT, Toffoli, and Fredkin gates,  with the same condition
	are shown in Figs. \ref{fig6} (b)-(d), respectively.
	If the amplitude $\alpha$ of the coherent state increases and dissipation coefficient $\gamma t$ decreases, the success probabilities of our schemes increase. By calculation with the different parameters of Ref. \cite{KerrCNOT1,Tgate-Dong}, six sets of data $P_{suc} (\alpha,\gamma t)$ are obtained, e.g.,
	$P_{suc}(40,1)=0.9282$, $P_{suc}(50,1)=0.9662$,
	$P_{suc}(40,0.5)=0.9698$, $P_{suc}(50,0.5)=0.9906$,
	$P_{suc}(40,0)=0.9921$, and $P_{suc}(50,0)=0.9987$.
	Based on the analyses above, the nearly deterministic quantum logical gates, i.e., CNOT,  Toffoli, and Fredkin gates in DFS, can be achieved.

	Above discussion about the success probabilities of our protocols, we only investigate the main factor $\alpha$.
	In view of the strength of natural cross-Kerr nonlinearity \cite{kerr-strength} is so small that it cannot provide effective interactions between photons, which leads to increasing the difficulty in distinguishing two overlapping coherent states and decreasing the success probabilities of the path coupler and three gates. Fortunately,
	various physical systems or synthetic media, such as negative-index metamaterials \cite{metamaterials}, a superconducting artificial atom \cite{0.35}, and a three-dimensional quantum electrodynamic architectures \cite{three-dimensional-1, three-dimensional-2}, have been explored to achieve stronger cross-Kerr nonlinearities. The paper in
	Ref. \cite{metamaterials}  mentioned that the Kerr nonlinearity $\chi$ ($10^{-4}$) can be achieved between the surface polaritons and the mutual phase shift was approximate $\pi$. Another paper in
	Ref. \cite{0.35} demonstrated the cross-Kerr phase shift up to $0.35$ ($\theta\simeq20^{\circ}$) based on per photon with the coherent microwave field at the single-photon level.
	Ref. \cite{three-dimensional-2} mentioned they designed phase shift large enough  ($\theta>30^{\circ}\approx0.52$).
	Figure \ref{fig7}a illustrates the success probability  $P_{suc} (\theta,\gamma t)$ 
	of the X-homodyne measurement of the path coupler vs the phase shift $\theta$
	with the different conditions $\gamma t=0$, $\gamma t=0.5$, and $\gamma t=1$ for fixed the
	parameters $\alpha=70$. If the phase shift $\theta$ increases and dissipation coefficient $\gamma t$ decreases, the success probabilities of our schemes increase.
	Figs. \ref{fig7} (b)-(d) show the corresponding success probabilities of our logic gates.
	By calculation with the different parameters of Ref. \cite{0.35,metamaterials,three-dimensional-1, three-dimensional-2}, six sets of data $P_{suc} (\theta,\gamma t)$ are obtained, e.g.,
	$P_{suc}(0.35,1)=0.9801$, $P_{suc}(0.52,1)=1$,
	$P_{suc}(0.35,0.5)=0.9981$, $P_{suc}(0.52,0.5)=1$,
	$P_{suc}(0.35,0)=1$, and $P_{suc}(0.52,0)=1$.
	Similarly, the success probabilities of three logical gates in DFS are close to 1.
	
	Besides,
	phase shift $\theta$ can be increased by prolonging the interaction time
	between photons and coherent state \cite{time1, time2, time3, time4},  as well as measurement-based methods \cite{measurement1, measurement2, measurement3} and quadrature squeezing operations \cite{quadrature} to achieve large phase shift.
	By employing a two-level atom in a one-sided cavity and the displacements-controlled photon number resolving (PNR) detector,  accomplished phase shift $\pi$ can be obtained \cite{2003}.
	Thus, in the case $\alpha\theta^{2}\gg1$, one can realize the deterministic distinguishability between the shifted and non-shifted phases in the coherent state, where the cross-Kerr nonlinearities $\theta\ll1$ with a sufficiently large amplitude of the coherent states can  be applicable.
	Furthermore,
	two coherent states can be discriminated to lower the error probability, resorting to a homodyne detector and a PNR detector applying the postselection strategy \cite{2010}.
	Therefore, we set up the CNOT, Toffoli and Fredkin gates in DFS based on Kerr effects, which can overcome and alleviate the disadvantageous factors, ultimately reducing the error probabilities.

	\subsection{Fidelities of our quantum gates regard to photon loss}

	The coupling between the coherent state and environment causes the loss of photons of the coherent state, so we consider the influence of photon loss on fidelity. The decoherence effects for the coherent state described by the standard master equation
	\begin{eqnarray}\label{eq26}
		\frac{\partial\rho}{\partial t}=\hat{J}\rho+\hat{L}\rho, \quad\hat{J}\rho=\gamma a\rho a^{\dag},\nonumber \\ \;\hat{L}\rho=-\frac{\gamma}{2}aa^{\dag}\rho+\rho aa^{\dag},
	\end{eqnarray}
	where $a$ and $a^{\dag}$ are the annihilation and creation operators of the coherent state, respectively. $\rho$ denotes the density matrix of the system.
	The formal solution of the master Eq. (\ref{eq26}) can be written as $\rho(t)=e^{(\hat{J}+\hat{L})t}\rho(0)$ and $t$ is the interaction time. The evolution of the hybrid
	system caused by nonlinear interactions between the photons and the coherent states could be described by a unitary evolution equation $\widetilde{U}(t)\rho(0)=U(t)\rho(0)U^{\dag}(t)$.
	Assumed that the interaction time $t$ is defined on the temporal interval $[a, b]$ and is divided into $N$ parts $a=x_{0}<x_{1}<x_{2}<...<x_{N}=b$ ($N\approx\infty$), setting $\Delta x_{N}=x_{N}-x_{N-1}$, the decoherence process $\widetilde{D}$ occurs for a short time $\Delta x_{N}$ and the unitary evolution operator $\widetilde{U}$ occurs for next temporal interval $\Delta x_{N+1}$ ($N=1, 2, 3...$).
	After the finite temporal interval $t$, the system would evolve as $\rho(t)=[\widetilde{D}(\Delta t)\widetilde{U}(\Delta t)]^{N}\rho(0)$ \cite{Dong-PRA}.

	As four scenarios of phase shifts, i.e., $0$, $\pm2\theta$, $\pm3\theta$, $\pm5\theta$ from the corresponding coherent states $|\alpha_{1}\rangle, |\alpha_{2}\rangle, |\alpha_{3}\rangle, $ and $|\alpha_{4}\rangle$ for implementation of the path coupler, two-logic-qubit CNOT gate, and three-logic-qubit Toffoli gate, and eight scenarios of phase shifts, i.e., $0, \pm\theta$, $\pm2\theta, \cdots$, $\pm7\theta$ from the coherent states $|\alpha_{5}\rangle$ for implementation of the three-logic-qubit Fredkin gate can be obtained by the X-homodyne measurements.
	We take the fidelity $F_{C}$ of the two-logic-qubit CNOT gate as an example to denote in detail. The initial state is $|\Omega\rangle$ in Eq. (\ref{eq9}), the density matrix of the system is $\rho(0)=|\Omega\rangle\langle\Omega|$.
	In the realm of Kerr media, the system is successively affected by $\widetilde{U}(\Delta t)$ as
	\begin{eqnarray}   \label{eq27}
		\widetilde{U}(\Delta t)\rho(0)&\!\!=\!\!&U(\Delta t)|\Omega\rangle\langle\Omega|U^{\dag}(\Delta t)  \nonumber\\
		&\!\!=\!\!&\sum_{k,l=-5}^{5}|\phi_{k}\rangle\langle\phi_{l}|\otimes|\alpha e^{i\frac{k\theta}{N}}\rangle\langle \alpha e^{i\frac{l\theta}{N}}|,\nonumber\\
		&&  k,l\in \{0, \pm2, \pm3, \pm5\}.
	\end{eqnarray}
	Here,
	\begin{eqnarray}   \label{eq}
		&&|\phi_{0}\rangle\;\,\,=
		\beta_{1}|\overline{0}\rangle_{AB}|\phi\rangle_{CD}|C_{2}\rangle|T_{1}T_{4}\rangle  \nonumber\\
		&&\qquad\quad+\beta_{2}|\overline{1}\rangle_{AB}|\phi\rangle_{CD}|C_{1}\rangle|T_{2}T_{3}\rangle,   \nonumber\\
		&&|\phi_{2}\rangle\;\,\,=\beta_{1}|\overline{0}\rangle_{AB}|\phi\rangle_{CD}|C_{2}\rangle|T_{1}T_{3}\rangle,  \nonumber\\
		&&|\phi_{-2}\rangle=\beta_{2}|\overline{1}\rangle_{AB}|\phi\rangle_{CD}|C_{1}\rangle|T_{2}T_{4}\rangle,  \nonumber\\
		&&|\phi_{3}\rangle\;\,\,=\beta_{1}|\overline{0}\rangle_{AB}|\phi\rangle_{CD}|C_{2}\rangle|T_{2}T_{4}\rangle,  \nonumber\\
		&&|\phi_{-3}\rangle=\beta_{2}|\overline{1}\rangle_{AB}|\phi\rangle_{CD}|C_{1}\rangle|T_{1}T_{3}\rangle,   \nonumber\\
		&&|\phi_{5}\rangle\;\,\,=\beta_{1}|\overline{0}\rangle_{AB}|\phi\rangle_{CD}|C_{2}\rangle|T_{2}T_{3}\rangle,  \nonumber\\
		&&|\phi_{-5}\rangle=\beta_{2}|\overline{1}\rangle_{AB}|\phi\rangle_{CD}|C_{1}\rangle|T_{1}T_{4}\rangle.
	\end{eqnarray}
	and $\widetilde{D}(\Delta t)$ as
	\begin{eqnarray}
		&&\widetilde{D}(\Delta t)\widetilde{U}(\Delta t)\rho(0)
		\nonumber\\
		&&=\sum_{k,l=-5}^{5}{\rm exp}\{\alpha^{2}(1-e^{-\gamma\frac{t}{N}})[e^{i(k-l)\frac{\theta}{N}}-1]\}\nonumber\\
		&&\times |\phi_{k}\rangle\langle\phi_{l}|\otimes|e^{-\frac{\gamma t}{2N}}\alpha e^{i\frac{k \theta}{N}}\rangle\langle e^{-\frac{\gamma t}{2N}}\alpha e^{i\frac{l \theta}{N}}|. 
	\end{eqnarray}
	Consequently, the system is changed into
	\begin{eqnarray}
		\rho(t)&\!\!=\!\!&[\widetilde{D}(\Delta t)\widetilde{U}(\Delta t)]^{N}\rho(0)    \nonumber\\
		&\!\!=\!\!&\sum_{k,l=-5}^{5}e^{B_{kl}}|\phi_{k}\rangle\langle\phi_{l}|\otimes|Ae^{ik\theta}\rangle\langle Ae^{il\theta}|, \nonumber\\
	\end{eqnarray}
	where $B_{kl}=\alpha^{2}(1-A^{\frac{2}{N}})\sum_{n=1}^{N}A^{\frac{2(n-1)}{N}}[e^{i(k-l)\frac{n\theta}{N}}-1]$.
	After performing X measurement on the dissipated coherent state, the resulting state can be obtained as follows
	\begin{widetext}
		\begin{eqnarray}
			\langle x |\rho(t)|x \rangle &=&\sum_{k,l=-5}^{5}C_{kl}|\phi_{k}\rangle\langle\phi_{l}|  \times f(x,k\theta)f(x,l\theta)e^{i[\delta_{kl}+\varphi(x,k\theta)-\varphi(x,l\theta)]},  
		\end{eqnarray}
	\end{widetext}
	where $C_{kl}=e^{{\rm Re}(B_{kl})}$, $\delta_{kl}={\rm Im}(B_{kl})$, the function $f(x,n\theta)$$=(2\pi)^{-\frac{1}{4}}{\rm exp}[-(x-2A\alpha\cos\theta)^{2}/4]$, and $\varphi(x,n\theta)$$=(x-2A\alpha\cos n\theta)A\alpha\sin n\theta$ ($n=0$, $\pm2$, $\pm3$, $\pm5$).
	
	\begin{figure}
		\centering
		\subfigure{
			\includegraphics[width=6.2cm]{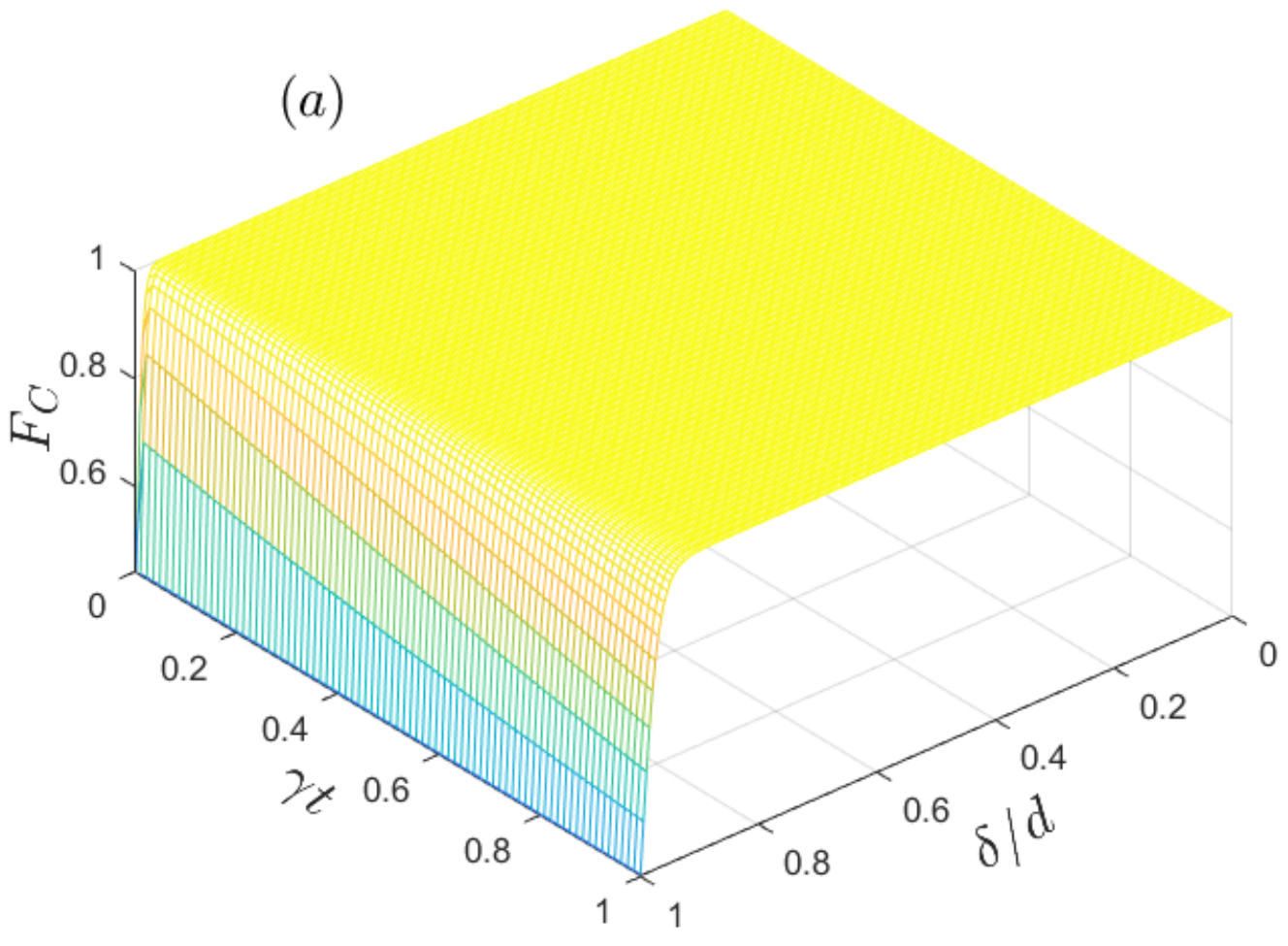}
		}
		\quad
		\subfigure{
			\includegraphics[width=6.2cm]{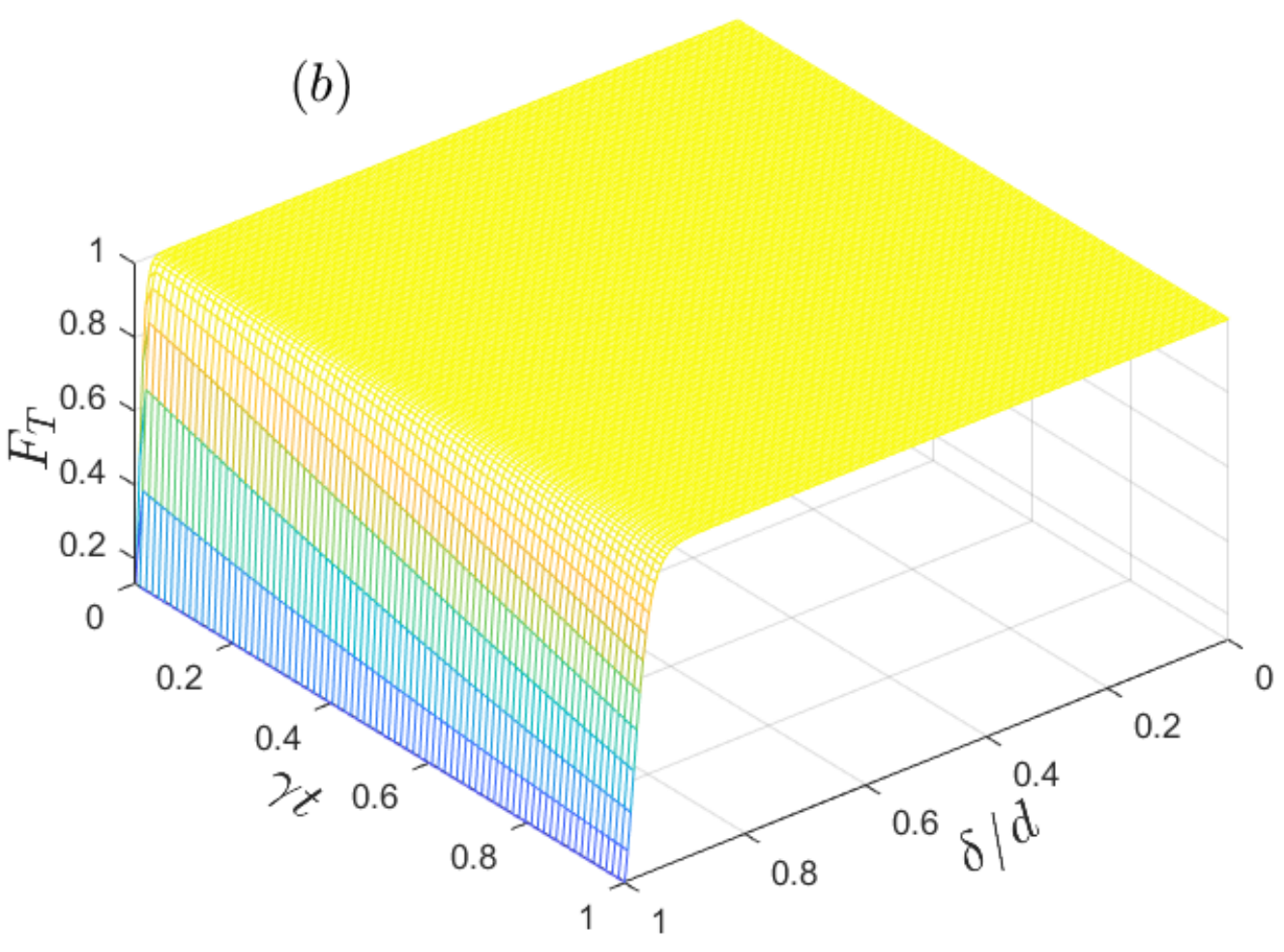}
		}
		\quad
		\subfigure{
			\includegraphics[width=6.2cm]{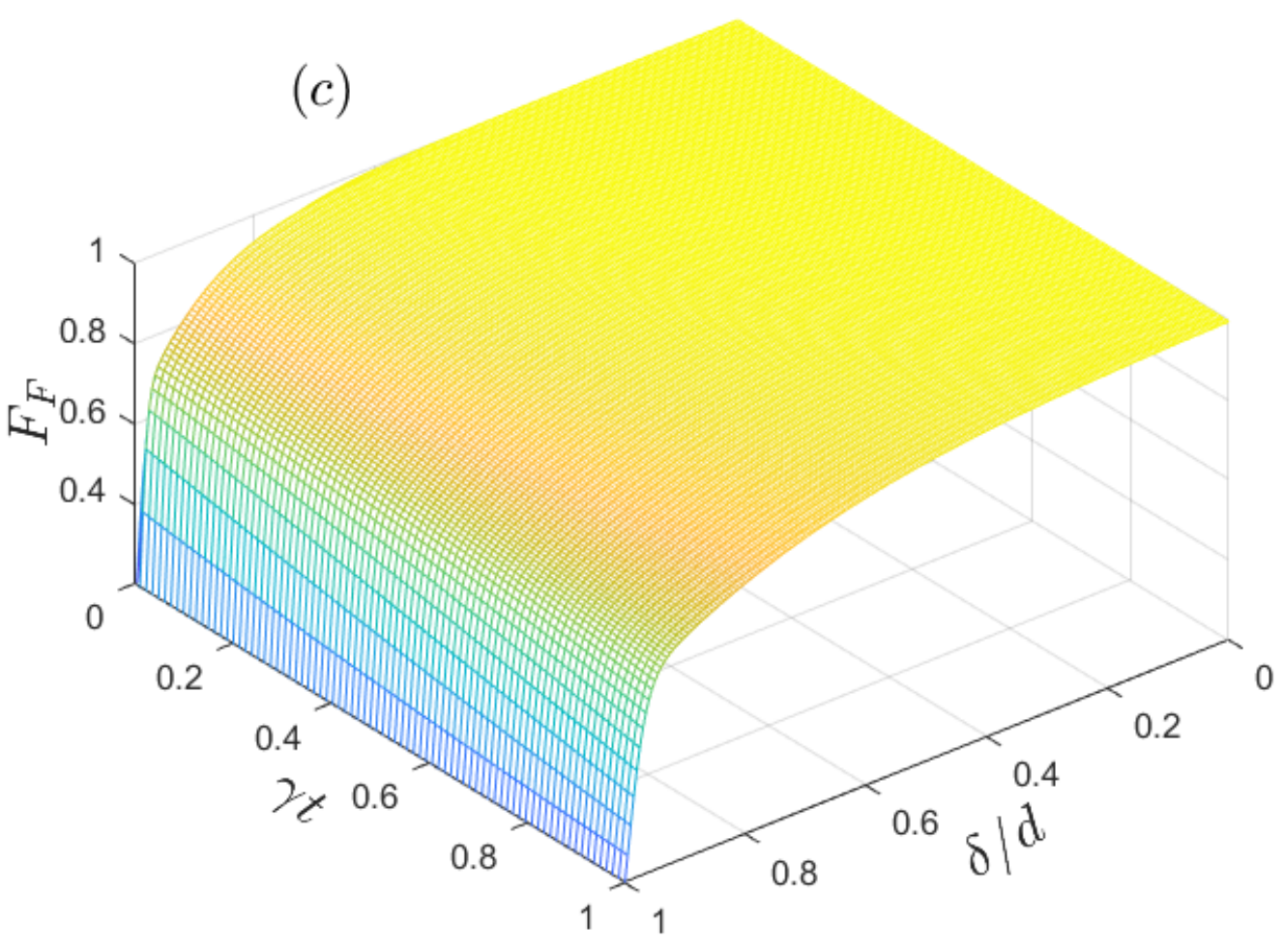}
		}
		\caption{  (a) The fidelity $F_{C}$ of CNOT gate (b) the fidelity $F_{F}$ of the Toffoli gate (c) the fidelity $F_{F}$ of the Fredkin gate vs the amplitude of the coherent state $\alpha$ and the dissipation coefficient $\gamma t$ for the parameters $N=200$, $\theta=20^{\circ}\approx0.35$ \cite{0.35}.}
		\label{fig8}
	\end{figure}

	Based on the fidelity formula $F=_{ideal}\langle\Psi|\rho_{out}|\Psi\rangle_{ideal}$  \cite{quadrature,Nielsen_Chuang_2010} determined by the real output state and the ideal one ($|\Omega\rangle_{3}, |\Phi\rangle_{5}$, or $|\Psi\rangle_{4}$), the fidelities of three gates vs the dissipation term $\gamma t$ as well as $\alpha$ with the parameters set as $N=200$ and $\theta=0.35$ \cite{0.35} are plotted in Figure \ref{fig8}.
	$X_{d_{n}}$ is the distance between the highest points of two adjacent Gaussian curves ($2d_{n}=X_{d_{n}}$), which increases leading to the reduction of overlap between the two Gauss functions.
	$\delta_{n}$ is the distance between measurement outcomes $x_{n}$ and the peak point ($x=2\alpha\cos n\theta$) of the Gaussian function curve $f(x,\alpha\cos n\theta)$.
	It can be seen that the fidelities $F_{C}$, $F_{T}$, and $F_{F}$ decrease
	with the increasing $\delta/d$ and
	the reducing  $\gamma t$.  By calculation with the different parameters of Ref. \cite{KerrCNOT1,Tgate-Dong}, four sets of data $F (\delta/d,\gamma t)$ are obtained, e.g.,
	$F_{C}(0.8,0)=1$, $F_{C}(0.8,1)=1$, $F_{C}(0.95,0)=0.9997$, $F_{C}(0.95,1)=0.9631$;
	$F_{T}(0.8,0)=1$, $F_{T}(0.8,1)=1$, $F_{T}(0.95,0)=0.9994$, $F_{T}(0.95,1)=0.9275$;
	$F_{F}(0.6,0)=0.9912$, $F_{F}(0.6,1)=0.9193$, $F_{F}(0.8,0)=0.914$, $F_{F}(0.8,1)=0.8268$.
	Based on the analyses above, the high fidelities of quantum logic gates in DFS can be achieved  even though under the decoherence environment.

	\section{\label{sec:5} SUMMARY}
	In summary, with the help of the cross-Kerr nonlinearities, we propose the effectuation of the nearly deterministic two-logic-qubit CNOT gate, three-logic-qubit Toffoli gate, and Fredkin gate encoded on polarization degrees of freedom of photon systems due to the significant features, e.g., the low-decoherence character, flexible single-qubit manipulation, and ultra-fast transmission.
	Moreover,
	the SWAP gates and the path couplers are innovated for implementing three logical gates.
	The SWAP gate, which serves to swap the polarized states of two photons as well as two
	paths of a single photon, is set up with simple linear-optics elements, so its success probability is 1,
	The path coupler with the probability near close to 1 is utilized to combine two paths of the photon to the one by available single-qubit operations and mature measurement methods, which halves the number of the paths of the photon and greatly simplifies the quantum circuits.
	Therefore,
	these three logical gates
	require neither complicated quantum computational circuits nor auxiliary photons (or entangled states).
	Further,
	the success probabilities of three logical gates are approximate 1
	by performing the corresponding classical feed-forward operations
	based on the different measuring results of the X-homodyne detectors to be aimed at the coherent states.
	Eventually,
	their fidelities are robust
	against the photon loss with the current technology.
	The proposed schemes play positive roles in the future development of QIP
	owing to the robust feature by resisting the influence of
	decoherence effect in DFS.
	\section*{ Acknowledgments}
	This work was supported in part by the Natural Science Foundation of China under Contract 61901420; 
	in part by Fundamental Research Program of Shanxi Province under Contract 20230302121116.
	\section*{ Disclosures}
	The authors declare that there are no conflicts of interest related to this article.
	Data Availability.
	\section*{ Data Availability}
	Data underlying the results presented in this paper are not publicly available at this time but may be obtained from the authors upon reasonable request.


\begin{thebibliography}{10}

\bibitem{Quantum2002}
Lov K.~Grover Michael A.~Nielsen, Isaac~Chuang.
\newblock ``Quantum computation and quantum information''.
\newblock \href{https://dx.doi.org/https://doi.org/10.1119/1.1463744}{Am. J.
  Phys. {\bf 70}, 558--559}~(2002).

\bibitem{Quan1}
Mingxia Huo and Ying Li.
\newblock ``Error-resilient {M}onte {C}arlo quantum simulation of imaginary
  time''.
\newblock \href{https://dx.doi.org/10.22331/q-2023-02-09-916}{{Quantum} {\bf
  7}, 916}~(2023).

\bibitem{Quan2}
Masahito Hayashi and Yuxiang Yang.
\newblock ``Efficient algorithms for quantum information bottleneck''.
\newblock \href{https://dx.doi.org/10.22331/q-2023-03-02-936}{{Quantum} {\bf
  7}, 936}~(2023).

\bibitem{Quan3}
XiuZhe Luo, JinGuo Liu, Pan Zhang, and Lei Wang.
\newblock ``Yao.jl: {E}xtensible, {E}fficient {F}ramework for {Q}uantum
  {A}lgorithm {D}esign''.
\newblock \href{https://dx.doi.org/10.22331/q-2020-10-11-341}{{Quantum} {\bf
  4}, 341}~(2020).

\bibitem{Quan4}
Xin Wang, Zhixin Song, and Youle Wang.
\newblock ``Variational {Q}uantum {S}ingular {V}alue {D}ecomposition''.
\newblock \href{https://dx.doi.org/10.22331/q-2021-06-29-483}{{Quantum} {\bf
  5}, 483}~(2021).

\bibitem{long2002theoretically}
G.~L. Long and X.~S. Liu.
\newblock ``Theoretically efficient high-capacity quantum-key-distribution
  scheme''.
\newblock \href{https://dx.doi.org/10.1103/PhysRevA.65.032302}{Phys. Rev. A
  {\bf 65}, 032302}~(2002).

\bibitem{zhang2017quantum}
Wei Zhang, Dong~Sheng Ding, Yu~Bo Sheng, Lan Zhou, Bao~Sen Shi, and Guang~Can
  Guo.
\newblock ``Quantum secure direct communication with quantum memory''.
\newblock Phys. Rev. Lett. {\bf 118}, 220501{\color{red}{ This bibitem caused a
  BibTeX warning: empty doi, eprint and url. If none of these fields is
  applicable for the cited work you can include the field }}
{\color{red}{\verb|`nolink = {}`| in its bib
  entry.}}\PackageError{quantum.bst}{The bibitem zhang2017quantum caused a
  BibTeX warning}{By default, quantum.bst handles BibTeX warnings like
  compilation errors. Please refer to the bibliography style demo for details.}

\bibitem{zhu2017experimental}
Feng Zhu, Wei Zhang, Yu~Bo Sheng, and Yi~Dong Huang.
\newblock ``Experimental long-distance quantum secure direct communication''.
\newblock Sci. Bull. {\bf 62}, 1519--1524{\color{red}{ This bibitem caused a
  BibTeX warning: empty doi, eprint and url. If none of these fields is
  applicable for the cited work you can include the field }}
{\color{red}{\verb|`nolink = {}`| in its bib
  entry.}}\PackageError{quantum.bst}{The bibitem zhu2017experimental caused a
  BibTeX warning}{By default, quantum.bst handles BibTeX warnings like
  compilation errors. Please refer to the bibliography style demo for details.}

\bibitem{du2019efficient}
Fang~Fang Du, Yong~Ting Liu, Zhen~Rong Shi, Yu~Xi Liang, Jun Tang, and Jun Liu.
\newblock ``Efficient hyperentanglement purification for three-photon systems
  with the fidelity-robust quantum gates and hyperentanglement link''.
\newblock Opt. Express {\bf 27}, 27046--27061{\color{red}{ This bibitem caused
  a BibTeX warning: empty doi, eprint and url. If none of these fields is
  applicable for the cited work you can include the field }}
{\color{red}{\verb|`nolink = {}`| in its bib
  entry.}}\PackageError{quantum.bst}{The bibitem du2019efficient caused a
  BibTeX warning}{By default, quantum.bst handles BibTeX warnings like
  compilation errors. Please refer to the bibliography style demo for details.}

\bibitem{li2020quantum}
Tao Li and Gui~Lu Long.
\newblock ``Quantum secure direct communication based on single-photon
  bell-state measurement''.
\newblock New J. Phys. {\bf 22}, 063017{\color{red}{ This bibitem caused a
  BibTeX warning: empty doi, eprint and url. If none of these fields is
  applicable for the cited work you can include the field }}
{\color{red}{\verb|`nolink = {}`| in its bib
  entry.}}\PackageError{quantum.bst}{The bibitem li2020quantum caused a BibTeX
  warning}{By default, quantum.bst handles BibTeX warnings like compilation
  errors. Please refer to the bibliography style demo for details.}

\bibitem{Quan5}
Zixin Huang, Peter~P. Rohde, Dominic~W. Berry, Pieter Kok, Jonathan~P. Dowling,
  and Cosmo Lupo.
\newblock ``Photonic quantum data locking''.
\newblock \href{https://dx.doi.org/10.22331/q-2021-04-28-447}{{Quantum} {\bf
  5}, 447}~(2021).

\bibitem{WANG202191}
Chuan Wang.
\newblock ``Quantum secure direct communication: Intersection of communication
  and cryptography''.
\newblock Fundam. Res. {\bf 1}, 91--92{\color{red}{ This bibitem caused a
  BibTeX warning: empty doi, eprint and url. If none of these fields is
  applicable for the cited work you can include the field }}
{\color{red}{\verb|`nolink = {}`| in its bib
  entry.}}\PackageError{quantum.bst}{The bibitem WANG202191 caused a BibTeX
  warning}{By default, quantum.bst handles BibTeX warnings like compilation
  errors. Please refer to the bibliography style demo for details.}

\bibitem{qi202115}
Zhantong Qi, Yuanhua Li, Yiwen Huang, Juan Feng, Yuanlin Zheng, and Xianfeng
  Chen.
\newblock ``A 15-user quantum secure direct communication network''.
\newblock Light Sci. Appl. {\bf 10}, 183{\color{red}{ This bibitem caused a
  BibTeX warning: empty doi, eprint and url. If none of these fields is
  applicable for the cited work you can include the field }}
{\color{red}{\verb|`nolink = {}`| in its bib
  entry.}}\PackageError{quantum.bst}{The bibitem qi202115 caused a BibTeX
  warning}{By default, quantum.bst handles BibTeX warnings like compilation
  errors. Please refer to the bibliography style demo for details.}

\bibitem{long2021drastic}
G.~L. Long and H.~Zhang.
\newblock ``Drastic increase of channel capacity in quantum secure direct
  communication using masking''.
\newblock Sci. Bull. {\bf 66}, 1267--1269{\color{red}{ This bibitem caused a
  BibTeX warning: empty doi, eprint and url. If none of these fields is
  applicable for the cited work you can include the field }}
{\color{red}{\verb|`nolink = {}`| in its bib
  entry.}}\PackageError{quantum.bst}{The bibitem long2021drastic caused a
  BibTeX warning}{By default, quantum.bst handles BibTeX warnings like
  compilation errors. Please refer to the bibliography style demo for details.}

\bibitem{sheng2022one}
Yu~Bo Sheng, Lan Zhou, and Gui~Lu Long.
\newblock ``One-step quantum secure direct communication''.
\newblock Sci. Bull. {\bf 67}, 367--374{\color{red}{ This bibitem caused a
  BibTeX warning: empty doi, eprint and url. If none of these fields is
  applicable for the cited work you can include the field }}
{\color{red}{\verb|`nolink = {}`| in its bib
  entry.}}\PackageError{quantum.bst}{The bibitem sheng2022one caused a BibTeX
  warning}{By default, quantum.bst handles BibTeX warnings like compilation
  errors. Please refer to the bibliography style demo for details.}

\bibitem{2024Cao-oe}
Yue~Ru Zhou, Qing~Feng Zhang, Fei~Fei Liu, Yu~Hong Han, Yong~Pan Gao, Ling Fan,
  Ru~Zhang, and Cong Cao.
\newblock ``Controllable nonreciprocal phonon laser in a hybrid photonic
  molecule based on directional quantum squeezing''.
\newblock Opt. Express {\bf 32}, 2786--2803{\color{red}{ This bibitem caused a
  BibTeX warning: empty doi, eprint and url. If none of these fields is
  applicable for the cited work you can include the field }}
{\color{red}{\verb|`nolink = {}`| in its bib
  entry.}}\PackageError{quantum.bst}{The bibitem 2024Cao-oe caused a BibTeX
  warning}{By default, quantum.bst handles BibTeX warnings like compilation
  errors. Please refer to the bibliography style demo for details.}

\bibitem{exper}
Peng Zhao, Meng~Ying Yang, Sha Zhu, Lan Zhou, Wei Zhong, Ming~Ming Du, and
  Yu~Bo Sheng.
\newblock ``Generation of hyperentangled state encoded in three degrees of
  freedom''.
\newblock \href{https://dx.doi.org/10.1007/s11433-023-2164-7}{Sci. China Phys.
  Mech. {\bf 66}, 100311}~(2023).

\bibitem{PhysRevA.105.032609}
Yun~Feng Guo, Wei Zhong, Lan Zhou, and Yu~Bo Sheng.
\newblock ``Supersensitivity of kerr phase estimation with two-mode squeezed
  vacuum states''.
\newblock Phys. Rev. A {\bf 105}, 032609{\color{red}{ This bibitem caused a
  BibTeX warning: empty doi, eprint and url. If none of these fields is
  applicable for the cited work you can include the field }}
{\color{red}{\verb|`nolink = {}`| in its bib
  entry.}}\PackageError{quantum.bst}{The bibitem PhysRevA.105.032609 caused a
  BibTeX warning}{By default, quantum.bst handles BibTeX warnings like
  compilation errors. Please refer to the bibliography style demo for details.}

\bibitem{cao2019high}
Cong Cao, Yu~Hong Han, Li~Zhang, Ling Fan, Yu~Wen Duan, and Ru~Zhang.
\newblock ``High-fidelity universal quantum controlled gates on electron-spin
  qubits in quantum dots inside single-sided optical microcavities''.
\newblock Adv. Quantum Technol. {\bf 2}, 1900081{\color{red}{ This bibitem
  caused a BibTeX warning: empty doi, eprint and url. If none of these fields
  is applicable for the cited work you can include the field }}
{\color{red}{\verb|`nolink = {}`| in its bib
  entry.}}\PackageError{quantum.bst}{The bibitem cao2019high caused a BibTeX
  warning}{By default, quantum.bst handles BibTeX warnings like compilation
  errors. Please refer to the bibliography style demo for details.}

\bibitem{CaoBELL}
Cong Cao, Li~Zhang, Yu~Hong Han, Pan~Pan Yin, Ling Fan, Yu~Wen Duan, and
  Ru~Zhang.
\newblock ``Complete and faithful hyperentangled-bell-state analysis of photon
  systems using a failure-heralded and fidelity-robust quantum gate''.
\newblock \href{https://dx.doi.org/10.1364/OE.384360}{Opt. Express {\bf 28},
  2857--2872}~(2020).

\bibitem{wu2022FOP}
Yi~Ming Wu, Gang Fan, and Fang~Fang Du.
\newblock ``Error-detected three-photon hyperparallel toffoli gate with
  state-selective reflection''.
\newblock Front. Phys. {\bf 17}, 51502{\color{red}{ This bibitem caused a
  BibTeX warning: empty doi, eprint and url. If none of these fields is
  applicable for the cited work you can include the field }}
{\color{red}{\verb|`nolink = {}`| in its bib
  entry.}}\PackageError{quantum.bst}{The bibitem wu2022FOP caused a BibTeX
  warning}{By default, quantum.bst handles BibTeX warnings like compilation
  errors. Please refer to the bibliography style demo for details.}

\bibitem{Zhang_2023}
Xiu~Yu Zhang, Cong Cao, Yong~Pan Gao, Ling Fan, Ru~Zhang, and Chuan Wang.
\newblock ``Generation and manipulation of phonon lasering in a two-drive
  cavity magnomechanical system''.
\newblock \href{https://dx.doi.org/10.1088/1367-2630/acd717}{New J. Phys. {\bf
  25}, 053039}~(2023).

\bibitem{Wei-npj}
Wen~Qiang Liu and Hai~Rui Wei.
\newblock ``Optimal synthesis of the fredkin gate in a multilevel system''.
\newblock New J. Phys. {\bf 22}, 063026{\color{red}{ This bibitem caused a
  BibTeX warning: empty doi, eprint and url. If none of these fields is
  applicable for the cited work you can include the field }}
{\color{red}{\verb|`nolink = {}`| in its bib
  entry.}}\PackageError{quantum.bst}{The bibitem Wei-npj caused a BibTeX
  warning}{By default, quantum.bst handles BibTeX warnings like compilation
  errors. Please refer to the bibliography style demo for details.}

\bibitem{Wei-PRApplied}
Wen~Qiang Liu, Hai~Rui Wei, and Leong~Chuan Kwek.
\newblock ``Low-cost fredkin gate with auxiliary space''.
\newblock Phys. Rev. Appl. {\bf 14}, 054057{\color{red}{ This bibitem caused a
  BibTeX warning: empty doi, eprint and url. If none of these fields is
  applicable for the cited work you can include the field }}
{\color{red}{\verb|`nolink = {}`| in its bib
  entry.}}\PackageError{quantum.bst}{The bibitem Wei-PRApplied caused a BibTeX
  warning}{By default, quantum.bst handles BibTeX warnings like compilation
  errors. Please refer to the bibliography style demo for details.}

\bibitem{17}
Jarom\'{\i}r Fiur\'a\ifmmode~\check{s}\else \v{s}\fi{}ek.
\newblock ``Linear optical fredkin gate based on partial-swap gate''.
\newblock Phys. Rev. A {\bf 78}, 032317{\color{red}{ This bibitem caused a
  BibTeX warning: empty doi, eprint and url. If none of these fields is
  applicable for the cited work you can include the field }}
{\color{red}{\verb|`nolink = {}`| in its bib
  entry.}}\PackageError{quantum.bst}{The bibitem 17 caused a BibTeX warning}{By
  default, quantum.bst handles BibTeX warnings like compilation errors. Please
  refer to the bibliography style demo for details.}

\bibitem{37}
Fabio Dell’Anno, Silvio {De Siena}, and Fabrizio Illuminati.
\newblock ``Multiphoton quantum optics and quantum state engineering''.
\newblock Phys. Rep. {\bf 428}, 53--168{\color{red}{ This bibitem caused a
  BibTeX warning: empty doi, eprint and url. If none of these fields is
  applicable for the cited work you can include the field }}
{\color{red}{\verb|`nolink = {}`| in its bib
  entry.}}\PackageError{quantum.bst}{The bibitem 37 caused a BibTeX warning}{By
  default, quantum.bst handles BibTeX warnings like compilation errors. Please
  refer to the bibliography style demo for details.}

\bibitem{wei2024PRApplied}
Gui-Long Jiang, Jun-Bin Yuan, Wen-Qiang Liu, and Hai-Rui Wei.
\newblock ``Efficient and deterministic high-dimensional controlled-swap gates
  on hybrid linear optical systems with high fidelity''.
\newblock Phys. Rev. Appl. {\bf 21}, 014001{\color{red}{ This bibitem caused a
  BibTeX warning: empty doi, eprint and url. If none of these fields is
  applicable for the cited work you can include the field }}
{\color{red}{\verb|`nolink = {}`| in its bib
  entry.}}\PackageError{quantum.bst}{The bibitem wei2024PRApplied caused a
  BibTeX warning}{By default, quantum.bst handles BibTeX warnings like
  compilation errors. Please refer to the bibliography style demo for details.}

\bibitem{KerrCNOT}
Kae Nemoto and W.~J. Munro.
\newblock ``Nearly deterministic linear optical controlled-not gate''.
\newblock Phys. Rev. Lett. {\bf 93}, 250502{\color{red}{ This bibitem caused a
  BibTeX warning: empty doi, eprint and url. If none of these fields is
  applicable for the cited work you can include the field }}
{\color{red}{\verb|`nolink = {}`| in its bib
  entry.}}\PackageError{quantum.bst}{The bibitem KerrCNOT caused a BibTeX
  warning}{By default, quantum.bst handles BibTeX warnings like compilation
  errors. Please refer to the bibliography style demo for details.}

\bibitem{KerrCNOT1}
Qing Lin and Jian Li.
\newblock ``Quantum control gates with weak cross-kerr nonlinearity''.
\newblock Phys. Rev. A {\bf 79}, 022301{\color{red}{ This bibitem caused a
  BibTeX warning: empty doi, eprint and url. If none of these fields is
  applicable for the cited work you can include the field }}
{\color{red}{\verb|`nolink = {}`| in its bib
  entry.}}\PackageError{quantum.bst}{The bibitem KerrCNOT1 caused a BibTeX
  warning}{By default, quantum.bst handles BibTeX warnings like compilation
  errors. Please refer to the bibliography style demo for details.}

\bibitem{KerrCNOT2}
Qing Lin and Bing He.
\newblock ``Single-photon logic gates using minimal resources''.
\newblock Phys. Rev. A {\bf 80}, 042310{\color{red}{ This bibitem caused a
  BibTeX warning: empty doi, eprint and url. If none of these fields is
  applicable for the cited work you can include the field }}
{\color{red}{\verb|`nolink = {}`| in its bib
  entry.}}\PackageError{quantum.bst}{The bibitem KerrCNOT2 caused a BibTeX
  warning}{By default, quantum.bst handles BibTeX warnings like compilation
  errors. Please refer to the bibliography style demo for details.}

\bibitem{Tgate-Dong}
Li~Dong, Sen~Lin Wang, Cen Cui, Xue Geng, Qing~Yang Li, Hai~Kuan Dong,
  Xiao~Ming Xiu, and Ya~Jun Gao.
\newblock ``Polarization toffoli gate assisted by multiple degrees of
  freedom''.
\newblock Opt. Lett. {\bf 43}, 4635--4638{\color{red}{ This bibitem caused a
  BibTeX warning: empty doi, eprint and url. If none of these fields is
  applicable for the cited work you can include the field }}
{\color{red}{\verb|`nolink = {}`| in its bib
  entry.}}\PackageError{quantum.bst}{The bibitem Tgate-Dong caused a BibTeX
  warning}{By default, quantum.bst handles BibTeX warnings like compilation
  errors. Please refer to the bibliography style demo for details.}

\bibitem{adv2}
Fang~Fang Du, Gang Fan, Xue~Mei Ren, and Ming Ma.
\newblock ``Deterministic hyperparallel control gates with weak kerr effects''.
\newblock \href{https://dx.doi.org/https://doi.org/10.1002/qute.202300201}{Adv.
  Quantum Technol. {\bf 6}, 2300201}~(2023).

\bibitem{weiNV}
Hai~Rui Wei and Fu~Guo Deng.
\newblock ``Compact quantum gates on electron-spin qubits assisted by diamond
  nitrogen-vacancy centers inside cavities''.
\newblock Phys. Rev. A {\bf 88}, 042323{\color{red}{ This bibitem caused a
  BibTeX warning: empty doi, eprint and url. If none of these fields is
  applicable for the cited work you can include the field }}
{\color{red}{\verb|`nolink = {}`| in its bib
  entry.}}\PackageError{quantum.bst}{The bibitem weiNV caused a BibTeX
  warning}{By default, quantum.bst handles BibTeX warnings like compilation
  errors. Please refer to the bibliography style demo for details.}

\bibitem{PRB-AiQing}
Qing Ai, Peng-Bo Li, Wei Qin, Jie-Xing Zhao, C.~P. Sun, and Franco Nori.
\newblock ``The nv metamaterial: Tunable quantum hyperbolic metamaterial using
  nitrogen vacancy centers in diamond''.
\newblock \href{https://dx.doi.org/10.1103/PhysRevB.104.014109}{Phys. Rev. B
  {\bf 104}, 014109}~(2021).

\bibitem{Rxm_OL}
Fang~Fang Du, Xue~Mei Ren, Ming Ma, and Gang Fan.
\newblock ``Qudit-based high-dimensional controlled-not gate''.
\newblock \href{https://dx.doi.org/10.1364/ol.518336}{Opt. Lett. {\bf 49},
  1229}~(2024).

\bibitem{WeiQD}
Hai~Rui Wei and Fu~Guo Deng.
\newblock ``Universal quantum gates on electron-spin qubits with quantum dots
  inside single-side optical microcavities''.
\newblock Opt. Express {\bf 22}, 593--607{\color{red}{ This bibitem caused a
  BibTeX warning: empty doi, eprint and url. If none of these fields is
  applicable for the cited work you can include the field }}
{\color{red}{\verb|`nolink = {}`| in its bib
  entry.}}\PackageError{quantum.bst}{The bibitem WeiQD caused a BibTeX
  warning}{By default, quantum.bst handles BibTeX warnings like compilation
  errors. Please refer to the bibliography style demo for details.}

\bibitem{wei-cpf}
Hai~Rui Wei, Yan~Bei Zheng, Ming Hua, and Guo~Fu Xu.
\newblock ``Robust-fidelity hyperparallel controlled-phase-flip gate through
  microcavities''.
\newblock Appl. Phys. Express {\bf 13}, 082007{\color{red}{ This bibitem caused
  a BibTeX warning: empty doi, eprint and url. If none of these fields is
  applicable for the cited work you can include the field }}
{\color{red}{\verb|`nolink = {}`| in its bib
  entry.}}\PackageError{quantum.bst}{The bibitem wei-cpf caused a BibTeX
  warning}{By default, quantum.bst handles BibTeX warnings like compilation
  errors. Please refer to the bibliography style demo for details.}

\bibitem{cao}
Yu~Hong Han, Cong Cao, Ling Fan, and Ru~Zhang.
\newblock ``Heralded high-fidelity quantum hyper-cnot gates assisted by charged
  quantum dots inside single-sided optical microcavities''.
\newblock Opt. Express {\bf 29}, 20045--20062{\color{red}{ This bibitem caused
  a BibTeX warning: empty doi, eprint and url. If none of these fields is
  applicable for the cited work you can include the field }}
{\color{red}{\verb|`nolink = {}`| in its bib
  entry.}}\PackageError{quantum.bst}{The bibitem cao caused a BibTeX
  warning}{By default, quantum.bst handles BibTeX warnings like compilation
  errors. Please refer to the bibliography style demo for details.}

\bibitem{SongPRA}
Guo~Zhu Song, Jin~Liang Guo, Qian Liu, Hai~Rui Wei, and Gui~Lu Long.
\newblock ``Heralded quantum gates for hybrid systems via waveguide-mediated
  photon scattering''.
\newblock Phys. Rev. A {\bf 104}, 012608{\color{red}{ This bibitem caused a
  BibTeX warning: empty doi, eprint and url. If none of these fields is
  applicable for the cited work you can include the field }}
{\color{red}{\verb|`nolink = {}`| in its bib
  entry.}}\PackageError{quantum.bst}{The bibitem SongPRA caused a BibTeX
  warning}{By default, quantum.bst handles BibTeX warnings like compilation
  errors. Please refer to the bibliography style demo for details.}

\bibitem{LiPRA1}
Tao Li, Adam Miranowicz, Ke~Yu Xia, and Franco Nori.
\newblock ``Resource-efficient analyzer of bell and greenberger-horne-zeilinger
  states of multiphoton systems''.
\newblock Phys. Rev. A {\bf 100}, 052302{\color{red}{ This bibitem caused a
  BibTeX warning: empty doi, eprint and url. If none of these fields is
  applicable for the cited work you can include the field }}
{\color{red}{\verb|`nolink = {}`| in its bib
  entry.}}\PackageError{quantum.bst}{The bibitem LiPRA1 caused a BibTeX
  warning}{By default, quantum.bst handles BibTeX warnings like compilation
  errors. Please refer to the bibliography style demo for details.}

\bibitem{SongPRA1}
Guo~Zhu Song, Ming~Jie Tao, Jing Qiu, and Hai~Rui Wei.
\newblock ``Quantum entanglement creation based on quantum scattering in
  one-dimensional waveguides''.
\newblock Phys. Rev. A {\bf 106}, 032416{\color{red}{ This bibitem caused a
  BibTeX warning: empty doi, eprint and url. If none of these fields is
  applicable for the cited work you can include the field }}
{\color{red}{\verb|`nolink = {}`| in its bib
  entry.}}\PackageError{quantum.bst}{The bibitem SongPRA1 caused a BibTeX
  warning}{By default, quantum.bst handles BibTeX warnings like compilation
  errors. Please refer to the bibliography style demo for details.}

\bibitem{DUcnot}
Fang~Fang Du and Zhen~Rong Shi.
\newblock ``Robust hybrid hyper-controlled-not gates assisted by an
  input-output process of low-q cavities''.
\newblock Opt. Express {\bf 27}, 17493--17506{\color{red}{ This bibitem caused
  a BibTeX warning: empty doi, eprint and url. If none of these fields is
  applicable for the cited work you can include the field }}
{\color{red}{\verb|`nolink = {}`| in its bib
  entry.}}\PackageError{quantum.bst}{The bibitem DUcnot caused a BibTeX
  warning}{By default, quantum.bst handles BibTeX warnings like compilation
  errors. Please refer to the bibliography style demo for details.}

\bibitem{adv1}
Fang~Fang Du, Yi~Ming Wu, and Gang Fan.
\newblock ``Refined quantum gates for Λ-type atom-photon hybrid systems''.
\newblock \href{https://dx.doi.org/https://doi.org/10.1002/qute.202300090}{Adv.
  Quantum Technol. {\bf 6}, 2300090}~(2023).

\bibitem{Google}
Frank Arute, Kunal Arya, Ryan Babbush, Dave Bacon, Joseph~C Bardin, Rami
  Barends, Rupak Biswas, Sergio Boixo, Fernando~GSL Brandao, David~A Buell,
  et~al.
\newblock ``Quantum supremacy using a programmable superconducting processor''.
\newblock Nature {\bf 574}, 505--510{\color{red}{ This bibitem caused a BibTeX
  warning: empty doi, eprint and url. If none of these fields is applicable for
  the cited work you can include the field }}
{\color{red}{\verb|`nolink = {}`| in its bib
  entry.}}\PackageError{quantum.bst}{The bibitem Google caused a BibTeX
  warning}{By default, quantum.bst handles BibTeX warnings like compilation
  errors. Please refer to the bibliography style demo for details.}

\bibitem{HV-encode}
Lu~Ming Duan and Guang~Can Guo.
\newblock ``Preserving coherence in quantum computation by pairing quantum
  bits''.
\newblock \href{https://dx.doi.org/10.1103/PhysRevLett.79.1953}{Phys. Rev.
  Lett. {\bf 79}, 1953--1956}~(1997).

\bibitem{QED-DFS}
Hua Wei, Wan~Li Yang, Zhi~Jiao Deng, and Mang Feng.
\newblock ``Many-qubit network employing cavity qed in a decoherence-free
  subspace''.
\newblock Phys. Rev. A {\bf 78}, 014304{\color{red}{ This bibitem caused a
  BibTeX warning: empty doi, eprint and url. If none of these fields is
  applicable for the cited work you can include the field }}
{\color{red}{\verb|`nolink = {}`| in its bib
  entry.}}\PackageError{quantum.bst}{The bibitem QED-DFS caused a BibTeX
  warning}{By default, quantum.bst handles BibTeX warnings like compilation
  errors. Please refer to the bibliography style demo for details.}

\bibitem{entangled-states}
Z.~J. Deng, M.~Feng, and K.~L. Gao.
\newblock ``Preparation of entangled states of four remote atomic qubits in
  decoherence-free subspace''.
\newblock Phys. Rev. A {\bf 75}, 024302{\color{red}{ This bibitem caused a
  BibTeX warning: empty doi, eprint and url. If none of these fields is
  applicable for the cited work you can include the field }}
{\color{red}{\verb|`nolink = {}`| in its bib
  entry.}}\PackageError{quantum.bst}{The bibitem entangled-states caused a
  BibTeX warning}{By default, quantum.bst handles BibTeX warnings like
  compilation errors. Please refer to the bibliography style demo for details.}

\bibitem{KLM}
Lei Chen, Xiao~Ming Xiu, Li~Dong, Shou Zhang, Shi~Lei Su, Shu Chen, and Er~Jun
  Liang.
\newblock ``Conversion of knill--laflamme--milburn entanglement to
  greenberger--horne--zeilinger entanglement in decoherence-free subspace''.
\newblock Ann. Phys. {\bf 534}, 2100365{\color{red}{ This bibitem caused a
  BibTeX warning: empty doi, eprint and url. If none of these fields is
  applicable for the cited work you can include the field }}
{\color{red}{\verb|`nolink = {}`| in its bib
  entry.}}\PackageError{quantum.bst}{The bibitem KLM caused a BibTeX
  warning}{By default, quantum.bst handles BibTeX warnings like compilation
  errors. Please refer to the bibliography style demo for details.}

\bibitem{GHZ}
Lei Chen, Xiao~Ming Xiu, Li~Dong, Nan~Nan Liu, Cai~Peng Shen, Shou Zhang, Shu
  Chen, and Shi~Lei Su.
\newblock ``Direct conversion of greenberger--horne--zeilinger state to
  knill--laflamme--milburn state in decoherence-free subspace''.
\newblock Opt. Lett. {\bf 47}, 2262--2265{\color{red}{ This bibitem caused a
  BibTeX warning: empty doi, eprint and url. If none of these fields is
  applicable for the cited work you can include the field }}
{\color{red}{\verb|`nolink = {}`| in its bib
  entry.}}\PackageError{quantum.bst}{The bibitem GHZ caused a BibTeX
  warning}{By default, quantum.bst handles BibTeX warnings like compilation
  errors. Please refer to the bibliography style demo for details.}

\bibitem{Rxm_OE}
Fang~Fang Du, Xue~Mei Ren, Zhi~Guo Fan, Ling~Hui Li, Xin~Shan Du, Ming Ma, Gang
  Fan, and Jing Guo.
\newblock ``Decoherence-free-subspace-based deterministic conversions for
  entangled states with heralded robust-fidelity quantum gates''.
\newblock \href{https://dx.doi.org/10.1364/oe.508088}{Opt. Express {\bf 32},
  1686}~(2024).

\bibitem{WOS:000766964200004}
Yi~Fan Qiao, Jia~Qiang Chen, Xing~Liang Dong, Bo~Long Wang, Xin~Lei Hei,
  Cai~Peng Shen, Yuan Zhou, and Peng~Bo Li.
\newblock ``Generation of greenberger-horne-zeilinger states for
  silicon-vacancy centers using a decoherence-free subspace''.
\newblock \href{https://dx.doi.org/10.1103/PhysRevA.105.032415}{Phys. Rev. A
  {\bf 105}, 032415}~(2022).

\bibitem{teleportation}
Hua Wei, ZhiJiao Deng, XiaoLong Zhang, and Mang Feng.
\newblock ``Transfer and teleportation of quantum states encoded in
  decoherence-free subspace''.
\newblock Phys. Rev. A {\bf 76}, 054304{\color{red}{ This bibitem caused a
  BibTeX warning: empty doi, eprint and url. If none of these fields is
  applicable for the cited work you can include the field }}
{\color{red}{\verb|`nolink = {}`| in its bib
  entry.}}\PackageError{quantum.bst}{The bibitem teleportation caused a BibTeX
  warning}{By default, quantum.bst handles BibTeX warnings like compilation
  errors. Please refer to the bibliography style demo for details.}

\bibitem{WOS:000501539900004}
Alejandro Fonseca.
\newblock ``High-dimensional quantum teleportation under noisy environments''.
\newblock \href{https://dx.doi.org/10.1103/PhysRevA.100.062311}{Phys. Rev. A
  {\bf 100}, 062311}~(2019).

\bibitem{PhysRevA.104.062612}
Xiao~Xiao Hu, Fei~Hao Zhang, Yan~Song Li, and Gui~Lu Long.
\newblock ``Optimizing quantum gates within decoherence-free subspaces''.
\newblock Phys. Rev. A {\bf 104}, 062612{\color{red}{ This bibitem caused a
  BibTeX warning: empty doi, eprint and url. If none of these fields is
  applicable for the cited work you can include the field }}
{\color{red}{\verb|`nolink = {}`| in its bib
  entry.}}\PackageError{quantum.bst}{The bibitem PhysRevA.104.062612 caused a
  BibTeX warning}{By default, quantum.bst handles BibTeX warnings like
  compilation errors. Please refer to the bibliography style demo for details.}

\bibitem{PhysRevLett.85.1758}
D.~Bacon, J.~Kempe, D.~A. Lidar, and K.~B. Whaley.
\newblock ``Universal fault-tolerant quantum computation on decoherence-free
  subspaces''.
\newblock \href{https://dx.doi.org/10.1103/PhysRevLett.85.1758}{Phys. Rev.
  Lett. {\bf 85}, 1758--1761}~(2000).

\bibitem{chen2010quantum}
Qiong Chen and Mang Feng.
\newblock ``Quantum-information processing in decoherence-free subspace with
  low-q cavities''.
\newblock Phys. Rev. A {\bf 82}, 052329{\color{red}{ This bibitem caused a
  BibTeX warning: empty doi, eprint and url. If none of these fields is
  applicable for the cited work you can include the field }}
{\color{red}{\verb|`nolink = {}`| in its bib
  entry.}}\PackageError{quantum.bst}{The bibitem chen2010quantum caused a
  BibTeX warning}{By default, quantum.bst handles BibTeX warnings like
  compilation errors. Please refer to the bibliography style demo for details.}

\bibitem{PhysRevLett.125.090501}
J.~S. Pedernales, F.~Cosco, and M.~B. Plenio.
\newblock ``Decoherence-free rotational degrees of freedom for quantum
  applications''.
\newblock Phys. Rev. Lett. {\bf 125}, 090501{\color{red}{ This bibitem caused a
  BibTeX warning: empty doi, eprint and url. If none of these fields is
  applicable for the cited work you can include the field }}
{\color{red}{\verb|`nolink = {}`| in its bib
  entry.}}\PackageError{quantum.bst}{The bibitem PhysRevLett.125.090501 caused
  a BibTeX warning}{By default, quantum.bst handles BibTeX warnings like
  compilation errors. Please refer to the bibliography style demo for details.}

\bibitem{hamann2022approximate}
Arne Hamann, Pavel Sekatski, and Wolfgang D{\"u}r.
\newblock ``Approximate decoherence free subspaces for distributed sensing''.
\newblock Quantum Sci. Technol. {\bf 7}, 025003{\color{red}{ This bibitem
  caused a BibTeX warning: empty doi, eprint and url. If none of these fields
  is applicable for the cited work you can include the field }}
{\color{red}{\verb|`nolink = {}`| in its bib
  entry.}}\PackageError{quantum.bst}{The bibitem hamann2022approximate caused a
  BibTeX warning}{By default, quantum.bst handles BibTeX warnings like
  compilation errors. Please refer to the bibliography style demo for details.}

\bibitem{parity-gate}
Xin~Wen Wang, Deng~Yu Zhang, Shi~Qing Tang, Li~Jun Xie, Zhi~Yong Wang, and
  Le~Man Kuang.
\newblock ``Photonic two-qubit parity gate with tiny cross--kerr
  nonlinearity''.
\newblock Phys. Rev. A {\bf 85}, 052326{\color{red}{ This bibitem caused a
  BibTeX warning: empty doi, eprint and url. If none of these fields is
  applicable for the cited work you can include the field }}
{\color{red}{\verb|`nolink = {}`| in its bib
  entry.}}\PackageError{quantum.bst}{The bibitem parity-gate caused a BibTeX
  warning}{By default, quantum.bst handles BibTeX warnings like compilation
  errors. Please refer to the bibliography style demo for details.}

\bibitem{0.35}
Io~Chun Hoi, Anton~F. Kockum, Tauno Palomaki, Thomas~M. Stace, Bixuan Fan, Lars
  Tornberg, Sankar~R. Sathyamoorthy, G\"oran Johansson, Per Delsing, and C.~M.
  Wilson.
\newblock ``Giant cross--kerr effect for propagating microwaves induced by an
  artificial atom''.
\newblock Phys. Rev. Lett. {\bf 111}, 053601{\color{red}{ This bibitem caused a
  BibTeX warning: empty doi, eprint and url. If none of these fields is
  applicable for the cited work you can include the field }}
{\color{red}{\verb|`nolink = {}`| in its bib
  entry.}}\PackageError{quantum.bst}{The bibitem 0.35 caused a BibTeX
  warning}{By default, quantum.bst handles BibTeX warnings like compilation
  errors. Please refer to the bibliography style demo for details.}

\bibitem{kerr-strength}
Pieter Kok, W.~J. Munro, Kae Nemoto, T.~C. Ralph, Jonathan~P. Dowling, and
  G.~J. Milburn.
\newblock ``Linear optical quantum computing with photonic qubits''.
\newblock Rev. Mod. Phys. {\bf 79}, 135--174{\color{red}{ This bibitem caused a
  BibTeX warning: empty doi, eprint and url. If none of these fields is
  applicable for the cited work you can include the field }}
{\color{red}{\verb|`nolink = {}`| in its bib
  entry.}}\PackageError{quantum.bst}{The bibitem kerr-strength caused a BibTeX
  warning}{By default, quantum.bst handles BibTeX warnings like compilation
  errors. Please refer to the bibliography style demo for details.}

\bibitem{metamaterials}
Michael Siomau, Ali~A. Kamli, Sergey~A. Moiseev, and Barry~C. Sanders.
\newblock ``Entanglement creation with negative index metamaterials''.
\newblock Phys. Rev. A {\bf 85}, 050303{\color{red}{ This bibitem caused a
  BibTeX warning: empty doi, eprint and url. If none of these fields is
  applicable for the cited work you can include the field }}
{\color{red}{\verb|`nolink = {}`| in its bib
  entry.}}\PackageError{quantum.bst}{The bibitem metamaterials caused a BibTeX
  warning}{By default, quantum.bst handles BibTeX warnings like compilation
  errors. Please refer to the bibliography style demo for details.}

\bibitem{three-dimensional-1}
Hanhee Paik, D.~I. Schuster, Lev~S. Bishop, G.~Kirchmair, G.~Catelani, A.~P.
  Sears, B.~R. Johnson, M.~J. Reagor, L.~Frunzio, L.~I. Glazman, S.~M. Girvin,
  M.~H. Devoret, and R.~J. Schoelkopf.
\newblock ``Observation of high coherence in josephson junction qubits measured
  in a three-dimensional circuit qed architecture''.
\newblock Phys. Rev. Lett. {\bf 107}, 240501{\color{red}{ This bibitem caused a
  BibTeX warning: empty doi, eprint and url. If none of these fields is
  applicable for the cited work you can include the field }}
{\color{red}{\verb|`nolink = {}`| in its bib
  entry.}}\PackageError{quantum.bst}{The bibitem three-dimensional-1 caused a
  BibTeX warning}{By default, quantum.bst handles BibTeX warnings like
  compilation errors. Please refer to the bibliography style demo for details.}

\bibitem{three-dimensional-2}
Gerhard Kirchmair, Brian Vlastakis, Zaki Leghtas, Simon~E. Nigg, Hanhee Paik,
  Eran Ginossar, Mazyar Mirrahimi, Luigi Frunzio, S.~M. Girvin, and R.~J.
  Schoelkopf.
\newblock ``Observation of quantum state collapse and revival due to the
  single-photon kerr effect''.
\newblock Nature {\bf 495}, 205--209{\color{red}{ This bibitem caused a BibTeX
  warning: empty doi, eprint and url. If none of these fields is applicable for
  the cited work you can include the field }}
{\color{red}{\verb|`nolink = {}`| in its bib
  entry.}}\PackageError{quantum.bst}{The bibitem three-dimensional-2 caused a
  BibTeX warning}{By default, quantum.bst handles BibTeX warnings like
  compilation errors. Please refer to the bibliography style demo for details.}

\bibitem{time1}
M.~D. Lukin and A.~Imamo\ifmmode~\breve{g}\else \u{g}\fi{}lu.
\newblock ``Nonlinear optics and quantum entanglement of ultraslow single
  photons''.
\newblock Phys. Rev. Lett. {\bf 84}, 1419--1422{\color{red}{ This bibitem
  caused a BibTeX warning: empty doi, eprint and url. If none of these fields
  is applicable for the cited work you can include the field }}
{\color{red}{\verb|`nolink = {}`| in its bib
  entry.}}\PackageError{quantum.bst}{The bibitem time1 caused a BibTeX
  warning}{By default, quantum.bst handles BibTeX warnings like compilation
  errors. Please refer to the bibliography style demo for details.}

\bibitem{time2}
M.~Bajcsy, A.~S. Zibrov, and M.~D. Lukin.
\newblock ``Stationary pulses of light in an atomic medium''.
\newblock Nature {\bf 426}, 638--641{\color{red}{ This bibitem caused a BibTeX
  warning: empty doi, eprint and url. If none of these fields is applicable for
  the cited work you can include the field }}
{\color{red}{\verb|`nolink = {}`| in its bib
  entry.}}\PackageError{quantum.bst}{The bibitem time2 caused a BibTeX
  warning}{By default, quantum.bst handles BibTeX warnings like compilation
  errors. Please refer to the bibliography style demo for details.}

\bibitem{time3}
Zeng~Bin Wang, Karl~Peter Marzlin, and Barry~C. Sanders.
\newblock ``Large cross-phase modulation between slow copropagating weak pulses
  in $^{87}\mathrm{Rb}$''.
\newblock Phys. Rev. Lett. {\bf 97}, 063901{\color{red}{ This bibitem caused a
  BibTeX warning: empty doi, eprint and url. If none of these fields is
  applicable for the cited work you can include the field }}
{\color{red}{\verb|`nolink = {}`| in its bib
  entry.}}\PackageError{quantum.bst}{The bibitem time3 caused a BibTeX
  warning}{By default, quantum.bst handles BibTeX warnings like compilation
  errors. Please refer to the bibliography style demo for details.}

\bibitem{time4}
Yi~Hsin Chen, Meng~Jung Lee, Weilun Hung, Ying~Cheng Chen, Yong~Fan Chen, and
  Ite~A. Yu.
\newblock ``Demonstration of the interaction between two stopped light
  pulses''.
\newblock Phys. Rev. Lett. {\bf 108}, 173603{\color{red}{ This bibitem caused a
  BibTeX warning: empty doi, eprint and url. If none of these fields is
  applicable for the cited work you can include the field }}
{\color{red}{\verb|`nolink = {}`| in its bib
  entry.}}\PackageError{quantum.bst}{The bibitem time4 caused a BibTeX
  warning}{By default, quantum.bst handles BibTeX warnings like compilation
  errors. Please refer to the bibliography style demo for details.}

\bibitem{measurement1}
Hyunseok Jeong.
\newblock ``Quantum computation using weak nonlinearities: Robustness against
  decoherence''.
\newblock Phys. Rev. A {\bf 73}, 052320{\color{red}{ This bibitem caused a
  BibTeX warning: empty doi, eprint and url. If none of these fields is
  applicable for the cited work you can include the field }}
{\color{red}{\verb|`nolink = {}`| in its bib
  entry.}}\PackageError{quantum.bst}{The bibitem measurement1 caused a BibTeX
  warning}{By default, quantum.bst handles BibTeX warnings like compilation
  errors. Please refer to the bibliography style demo for details.}

\bibitem{measurement2}
Amir Feizpour, Xingxing Xing, and Aephraim~M. Steinberg.
\newblock ``Amplifying single-photon nonlinearity using weak measurements''.
\newblock Phys. Rev. Lett. {\bf 107}, 133603{\color{red}{ This bibitem caused a
  BibTeX warning: empty doi, eprint and url. If none of these fields is
  applicable for the cited work you can include the field }}
{\color{red}{\verb|`nolink = {}`| in its bib
  entry.}}\PackageError{quantum.bst}{The bibitem measurement2 caused a BibTeX
  warning}{By default, quantum.bst handles BibTeX warnings like compilation
  errors. Please refer to the bibliography style demo for details.}

\bibitem{measurement3}
Seckin Sefi, Vishal Vaibhav, and Peter van Loock.
\newblock ``Measurement induced optical kerr interaction''.
\newblock Phys. Rev. A {\bf 88}, 012303{\color{red}{ This bibitem caused a
  BibTeX warning: empty doi, eprint and url. If none of these fields is
  applicable for the cited work you can include the field }}
{\color{red}{\verb|`nolink = {}`| in its bib
  entry.}}\PackageError{quantum.bst}{The bibitem measurement3 caused a BibTeX
  warning}{By default, quantum.bst handles BibTeX warnings like compilation
  errors. Please refer to the bibliography style demo for details.}

\bibitem{quadrature}
Monika Bartkowiak, Lian~Ao Wu, and Adam Miranowicz.
\newblock ``Quantum circuits for amplification of kerr nonlinearity via
  quadrature squeezing''.
\newblock Journal of Physics B: Atomic, Molecular and Optical Physics {\bf 47},
  145501{\color{red}{ This bibitem caused a BibTeX warning: empty doi, eprint
  and url. If none of these fields is applicable for the cited work you can
  include the field }}
{\color{red}{\verb|`nolink = {}`| in its bib
  entry.}}\PackageError{quantum.bst}{The bibitem quadrature caused a BibTeX
  warning}{By default, quantum.bst handles BibTeX warnings like compilation
  errors. Please refer to the bibliography style demo for details.}

\bibitem{2003}
Holger~F Hofmann, Kunihiro Kojima, Shigeki Takeuchi, and Keiji Sasaki.
\newblock ``Optimized phase switching using a single-atom nonlinearity''.
\newblock J. Opt. B: Quantum Semiclass. Opt. {\bf 5}, 218{\color{red}{ This
  bibitem caused a BibTeX warning: empty doi, eprint and url. If none of these
  fields is applicable for the cited work you can include the field }}
{\color{red}{\verb|`nolink = {}`| in its bib
  entry.}}\PackageError{quantum.bst}{The bibitem 2003 caused a BibTeX
  warning}{By default, quantum.bst handles BibTeX warnings like compilation
  errors. Please refer to the bibliography style demo for details.}

\bibitem{2010}
Christoffer Wittmann, Ulrik~L. Andersen, Masahiro Takeoka, Denis Sych, and Gerd
  Leuchs.
\newblock ``Discrimination of binary coherent states using a homodyne detector
  and a photon number resolving detector''.
\newblock Phys. Rev. A {\bf 81}, 062338{\color{red}{ This bibitem caused a
  BibTeX warning: empty doi, eprint and url. If none of these fields is
  applicable for the cited work you can include the field }}
{\color{red}{\verb|`nolink = {}`| in its bib
  entry.}}\PackageError{quantum.bst}{The bibitem 2010 caused a BibTeX
  warning}{By default, quantum.bst handles BibTeX warnings like compilation
  errors. Please refer to the bibliography style demo for details.}

\bibitem{Dong-PRA}
Li~Dong, Jun~Xi Wang, Qing~Yang Li, Hong~Zhi Shen, Hai~Kuan Dong, Xiao~Ming
  Xiu, Ya~Jun Gao, and Choo~Hiap Oh.
\newblock ``Nearly deterministic preparation of the perfect $w$ state with weak
  cross-kerr nonlinearities''.
\newblock \href{https://dx.doi.org/10.1103/PhysRevA.93.012308}{Phys. Rev. A
  {\bf 93}, 012308}~(2016).

\bibitem{Nielsen_Chuang_2010}
Michael~A. Nielsen and Isaac~L. Chuang.
\newblock ``Quantum computation and quantum information''.
\newblock Cambridge University Press. {\color{red}{ This bibitem caused a
  BibTeX warning: empty doi, eprint and url. If none of these fields is
  applicable for the cited work you can include the field }}
{\color{red}{\verb|`nolink = {}`| in its bib
  entry.}}\PackageError{quantum.bst}{The bibitem Nielsen_Chuang_2010 caused a
  BibTeX warning}{By default, quantum.bst handles BibTeX warnings like
  compilation errors. Please refer to the bibliography style demo for details.}

\end{thebibliography}

\end{document}